\documentclass{article}

\usepackage{arxiv}

\usepackage[colorlinks=true, allcolors=blue]{hyperref}
\usepackage[utf8]{inputenc} % allow utf-8 input
\usepackage[T1]{fontenc}    % use 8-bit T1 fonts
\usepackage{url}            % simple URL typesetting
\usepackage{booktabs}       % professional-quality tables
\usepackage{amsfonts}       % blackboard math symbols
\usepackage{nicefrac}       % compact symbols for 1/2, etc.
\usepackage{microtype}      % microtypography
\usepackage{lipsum}         % Can be removed after putting your text content
\usepackage{graphicx}
\usepackage{natbib}
\usepackage{doi}
\usepackage{ulem}
\usepackage{tikz}
\usepackage{marginnote}
\usepackage{soul}
\usepackage{xcolor}
\definecolor{emerald}{HTML}{50C878}
\usetikzlibrary{arrows.meta, calc}

\usepackage[english]{babel}

\usepackage{amsmath}
\usepackage{graphicx}
\usepackage{makecell}

\usepackage{tikz,marginnote}
\usepackage{ifoddpage}

\usepackage{cleveref}
\crefrangelabelformat{equation}{(#1-#2)}

\newcounter{todocounter}

\title{AICON: An operational global machine learning weather forecasting model}

\author{Tobias Göcke \\ Deutscher Wetterdienst  \\ tobias.goecke@dwd.de
   \And Marek Jacob \\ Deutscher Wetterdienst 
   \And Florian Prill \\ Deutscher Wetterdienst   
   \AND 
        Michael Denhard \\ Deutscher Wetterdienst
  \And Felix Fundel \\ Deutscher Wetterdienst
   \And Jan Keller \\ Deutscher Wetterdienst
   \And Roland Potthast \\ Deutscher Wetterdienst \\ University of Reading, UK
   \And Hendrik Reich \\ Deutscher Wetterdienst
    \And 
        Britta Seegebrecht\\
        Deutscher Wetterdienst, \\ LMU  Munich
   \And Sven Ulbrich \\ Deutscher Wetterdienst
    \And Arianna Valmassoi \\ Deutscher Wetterdienst
    \And Sabrina Wahl \\ Deutscher Wetterdienst
        }

\begin{document}
\maketitle

\begin{abstract}

We introduce AICON, a global machine learning weather prediction (MLWP) model which generates forecasts at 13 km spatial resolution with a 3-hour time step, trained on the high-resolution, non-hydrostatic ICON-DREAM dataset. AICON is in full operational use at Deutscher Wetterdienst since 2nd of March 2026. The model employs a graph neural network (GNN) architecture with an encoder-processor-decoder structure, where node updates are performed using a graph attention mechanism. A key feature of AICON is its use of an icosahedral multi-mesh derived from the native grid of the ICON model, ensuring consistency with the training data. ICON's terrain-following vertical SLEVE coordinate is one of the major distinctions from existing emulators.
AICON's training strategy prioritizes small-scale fidelity by avoiding autoregressive multi-step rollout and longer forecast horizons during training, a design choice motivated by the hypothesis that this approach preserves fine-scale features often damped in models optimized for longer-range forecasts. We describe the prognostic and diagnostic variables used for training, the transfer learning protocol employed to accelerate convergence, and the model’s performance across a range of evaluation metrics. An extensive evaluation, including routine verification against observation, a tropical cyclone case and spectral analysis reveal the strengths and limitations in the representation of atmospheric variability across scales. Routine verification against observations demonstrates competitive skill relative to the operational ICON model, particularly for near-surface variables in the short to medium forecast range. 
\end{abstract}

\section{Introduction}\label{sec:introduction}

The rapid advancement of machine learning weather prediction (MLWP) models since 2022 has fundamentally transformed the landscape of numerical weather prediction (NWP). Recent MLWP models 
(e.g. \cite{pathak2022fourcastnet,bi2022pangu,lam2023learning,lang2024aifsecmwfsdatadriven} ) have demonstrated forecasting capabilities that rival or even surpass conventional physics-based systems in certain aspects \cite{benbouallegue2023risedatadrivenweatherforecasting,bauer2024if}. A key advantage of MLWP models lies in their computational efficiency, enabling applications like large ensemble forecasts that were previously infeasible with traditional NWP systems. While the future trajectory of NWP remains uncertain in light of these developments, it is clear that weather centers must adapt by integrating MLWP models into their forecasting suites \cite{RABIER2026100040}. \\

In this context, the German Meteorological Service (Deutscher Wetterdienst, DWD) has taken a strategic step forward by developing AICON global (hereafter referred to as AICON), a model trained on the global ICON-DREAM reanalysis dataset \cite{valmassoi2025}.
A strong emphasis has been put on an intensive development and review process, with technical operations since September 2025 and full operational use since 2nd of March 2026 \cite{dwd2026aicon}. This also facilitated an extensive verification within a daily forecast routine by DWDs forecasting department. Unlike most existing global MLWP models, which rely on the ERA5 reanalysis \cite{hersbach2020era5}, AICON is trained on the ICON-DREAM dataset, which is based on DWD’s operational global forecasting system (as of March 2024). This dataset offers distinct advantages, including a higher spatial resolution (13 km) and temporal frequency (3-hourly) compared to ERA5, as well as a non-hydrostatic dynamical core and a state-of-the-art ensemble variational data assimilation scheme \cite{zaengl2015,prill2025,PotthastEtAl2019DACE}. These properties are expected to enhance the representation of small-scale atmospheric features and improve forecast fidelity, particularly for near-surface variables where the ICON model has demonstrated exceptional skill \cite{Zaengl_2023}. \\

The model employs a graph neural network (GNN) architecture with an encoder–processor–decoder structure \cite{lam2023learning}, where node updates are performed using a graph attention mechanism \cite{shi2021}. A defining feature of AICON is its use of an icosahedral multi-mesh derived from the native grid of the ICON model, facilitating a successive increase of
the model resolution during training by applying an efficient transfer learning protocol described in Section~\ref{sec:transfer_learning}. 

A central concept guiding the development of AICON is the preservation of small scale model fidelity. After initially following the standard approach and training using
autoregressive multi-step rollout in training we abandoned this training strategy. Most MLWP models use longer forecast horizons in the training
process to improve the headline scores and long-term stability during inference \cite{bi2022pangu,bodnar2025foundation,lam2023learning,lang2024aifsecmwfsdatadriven}. 
Either long forecast steps (up to 24 hours or more are) are used as basic model step, or the model output is fed back again auto-regressively several times and a multi-step loss function is applied.
We experiment with a eight-fold 3~h step resulting in 24 h forecast horizon in training but find that this
is in contradiction with small scale activity. It is by now well known that deterministic MLWP models tend to suppress spatial scales that are unpredictable beyond short time horizons (e.g., 3–6 hours) \cite{selz2025effective,selz2023can} if multi-step rollout is applied. 
We come back to this in the evaluation section \ref{sec:rollout_vs_no_rollout}. 
Further, AICON is implemented within the Anemoi framework \cite{aifsblogAnemoi,Wijnands2025Anemoi}, a collaborative platform developed by ECMWF and European National Meteorological Services (NMS) to facilitate the development of data-driven weather models.  \\ 

The remainder of this paper is organized as follows: In Section \ref{sec:icon-dream}, we describe the ICON-DREAM reanalysis dataset, highlighting its unique properties and the implications for training AICON. Section \ref{sec:model} details the AICON model architecture, including its graph neural network design, the selection of prognostic and diagnostic variables, and the training strategy. 
The details of the training procedure are presented in section \ref{sec:training}. After a short description of the operationalization in section \ref{sec:oper}, the evaluation of AICON is presented in Section \ref{sec:evaluation}, where we assess its performance through routine verification against observations, a case study of Hurricane Sinlaku, and a spectral analysis of forecast variability. Finally, Section \ref{sec:summary} summarizes our findings, discusses the implications of AICON’s operational implementation at DWD, and outlines directions for future work.

\section{Training data: ICON-DREAM reanalysis}\label{sec:icon-dream}

Unlike the majority of MLWP models, which train on the ERA5 reanalysis~\cite{hersbach2020era5}, 
AICON is trained on the ICON-DREAM reanalysis~\cite{valmassoi2025}, based on DWD's operational 
forecasting system described in \cite{zaengl2015,prill2025}, see also \cite{PotthastWalterRhodin2019LAPF,SchenkPotthastRojahn2022LocalizedPF}. This constitutes a distinct data environment in terms 
of the underlying NWP model, data assimilation scheme, and output representation, with implications 
for model design and skill that are explored in Section~\ref{sec:evaluation}.

\paragraph{ICON dynamical core.}
ICON-DREAM is generated with the ICOsahedral Nonhydrostatic (ICON) modeling 
framework~\cite{zaengl2015,prill2025, PotthastEtAl2019DACE}, configured as of March 2024. ICON solves the fully 
compressible, non-hydrostatic equations of atmospheric motion on an unstructured triangular 
grid. Its prognostic base state comprises the horizontal wind $\mathbf{v}_h$, vertical wind $w$,
Exner pressure $\pi$, total density $\overline{\rho}$ and partial mass fractions $\hat{q}_k$ for water
vapor and hydrometers supplemented by physical parameterizations covering turbulence, convection, 
microphysics, and radiation. Vertically, ICON employs terrain-following SLEVE 
coordinates~\cite{schaer2002sleve}, and output is stored on native model levels. AICON is 
consequently trained on model levels rather than on pressure levels, preserving the 
native vertical discretization of the analysis.

\paragraph{ICON-DREAM configuration and data assimilation.}
The reanalysis is run at 13~km resolution globally with a 3-hourly data assimilation cycle. The assimilation scheme is an Ensemble Variational (EnVar, see \cite{PotthastEtAl2019DACE, prill2025}) method, with flow-dependent background error covariances 
supplied by a 20-member ensemble at 40~km using an LETKF 
scheme~\cite{hunt2007efficient}. Complementary analyses include snow updates every 3~h, 
a daily soil moisture analysis at 00~UTC, and sea surface temperatures from the 
$1/20\text{\textdegree}$ OSTIA product~\cite{donlon2012operational}. This is different from the 4DVar used for ERA5. We note that the configuration of the reanalysis is slightly modified from the operational system to achieve higher throughput, with nearly operational quality. 

\paragraph{Implications and expectations.}
The most notable differences of ICON-DREAM compared to ERA5 are the spatial and temporal 
resolutions (13 km vs $\sim$ 31 km and 3 hours vs 6 hours). Whereas the mesh size obviously 
holds the promise for an increase in small scale activity, the same is true for the smaller
time step the model can be trained on. This is because deterministic MLWP models dampen
spatial scales that are unpredictable for a given time horizon (e.g. 3 hours or 6 hours if no autoregressive multi-step rollout is applied in training).
So after 6 hours the smallest predictable scales are coarser than after 3 hours, see also \cite{selz2025effective}.
One of the central goals is thus to achieve as much small scale fidelity as possible given the training data set
at hand. This means also, as will be detailed in section \ref{sec:training_curriculum}, that we will
refrain from using multi step rollout in training.
The other differences compared to ERA5, the vertical levels, non-hydrostatic equations and different data assimilation scheme
are further arguments why training on ICON-DREAM is an interesting alternative, but their impact on the final MLWP model is less obvious.

\section{The model}\label{sec:model}

In this chapter we describe the AICON model, starting with 
an overview over the model architecture in section \ref{sec:architecture},
followed by a short description
of the  underlying software framework Anemoi in section \ref{sec:anemoi-framework}

\subsection{Model architecture}\label{sec:architecture}

Deterministic weather forecasting can be formulated in an abstract way as
\begin{align*}
\bigl[ x_{t-(\alpha-1)}, \ldots, x_t \bigr] \overset{\mathcal{F}(\theta)}{\rightarrow} \bigl[ y_{t+1}, \ldots, y_{t+\beta} \bigr]
\end{align*}
where $x$ and $y$ are sets of input and output variables (feature vector); $\alpha$ and $\beta$ are the temporal lengths of the input and output windows and $\mathcal{F}(\theta)$ represents the model with the learnable parameters $\theta$. For AICON, the model receives as input a representation of the atmospheric states at $t_{-3\text{h}}$, $t_0$ and then forecasts the state at time $t_{+3\text{h}}$, i.e. we have $\alpha := 2$, $\beta := 1$. Forecasts for longer lead-times are calculated by initializing the model from its own prediction. This method is known as autoregressive rollout.

Inspired by Keisler's work on Graph Neural Networks (GNNs) for numerical weather prediction~\cite{keisler2022forecasting}, the AICON model has three main building blocks, which are provided by the general encoder-processor-decoder architecture of the Anemoi framework and are in common with the AIFS model~\cite{lang2024aifsecmwfsdatadriven}: First, an encoder transforms raw node and edge features into latent representations suitable for graph processing. Then the processor iteratively updates node embeddings by aggregating and exchanging information across the graph structure using message passing. Finally, the decoder maps the final node or graph embeddings to the desired output. In general, the model uses skip connections, which add the input directly to the output of the model layers. This makes the network focus on learning increments.

\paragraph{Graph structure.}

\begin{figure}
    \centering
    % AICON HIDDEN MESH ILLUSTRATION
    % "Hidden mesh of AICON. R03B03 example."
    % 
    % created by: F. Prill, DWD
    % creation date: 2026-07-13
    % tools: Python (Jupyter notebook)
    %        plot_aicon_grid/plot_hidden01.ipynb
    \includegraphics[width=0.5\linewidth]{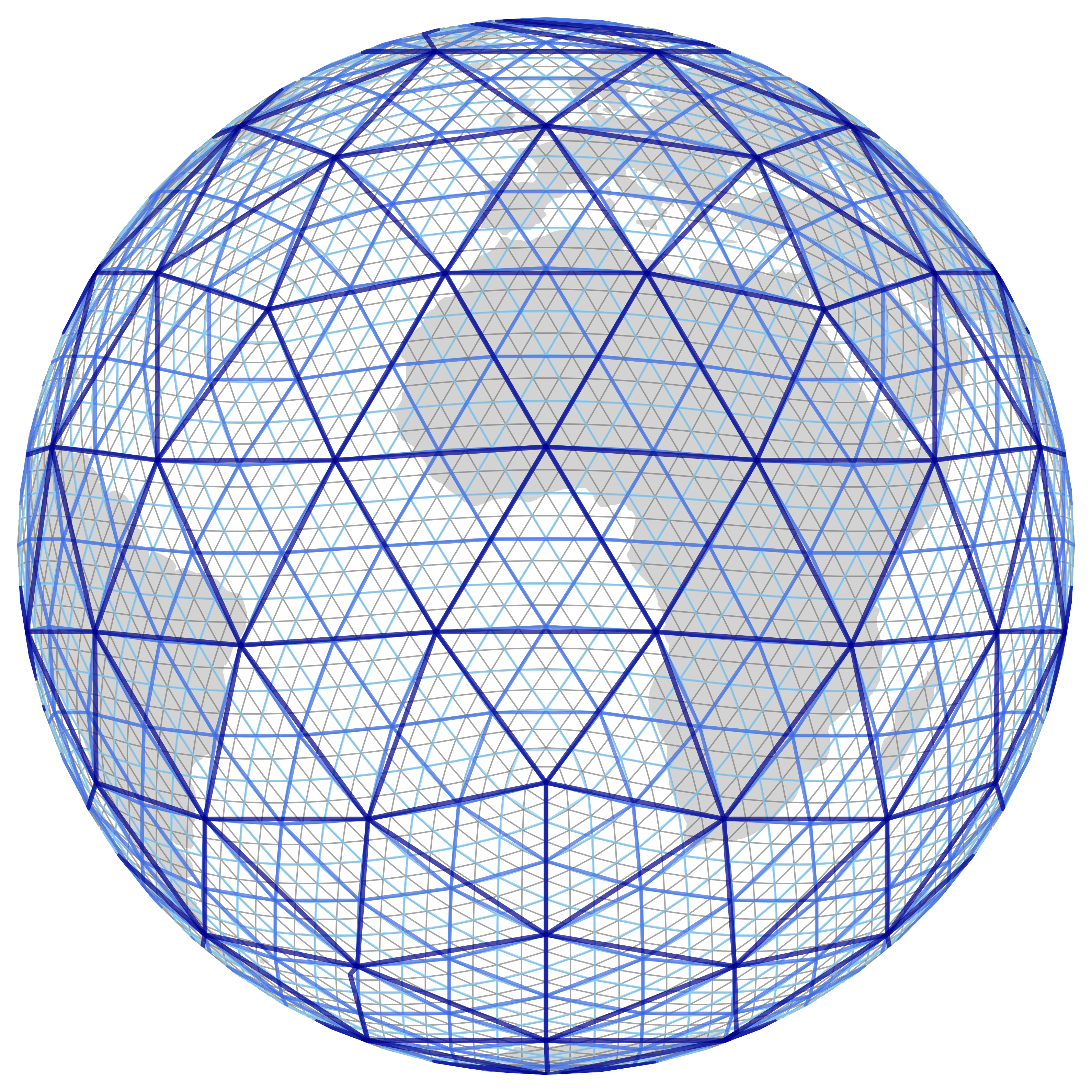}
    \caption{Hidden mesh of AICON. R03B03 example. The set of edges in the hidden mesh is equivalent to the union of all edges in the coarse-to-fine level hierarchy $RnB0, \ldots, RnBk$.}
    \label{fig:hidden_mesh_aicon}
\end{figure}

\begin{figure}
    \centering
    % AICON encoder/decoder graph visualization
    % "Examples of local graph structure (R03B04). Blue: a hidden node (red) connected to 24 data nodes (black dots) in the encoder. Green: Data node connected to its surrounding three hidden nodes in the decoder."
    % 
    % created by: F. Prill, DWD
    % creation date: 2026-07-13
    % tools: Python (Jupyter notebook)
    %        plot_aicon_grid/plot_hidden01.ipynb
    \includegraphics[width=1.0\linewidth]{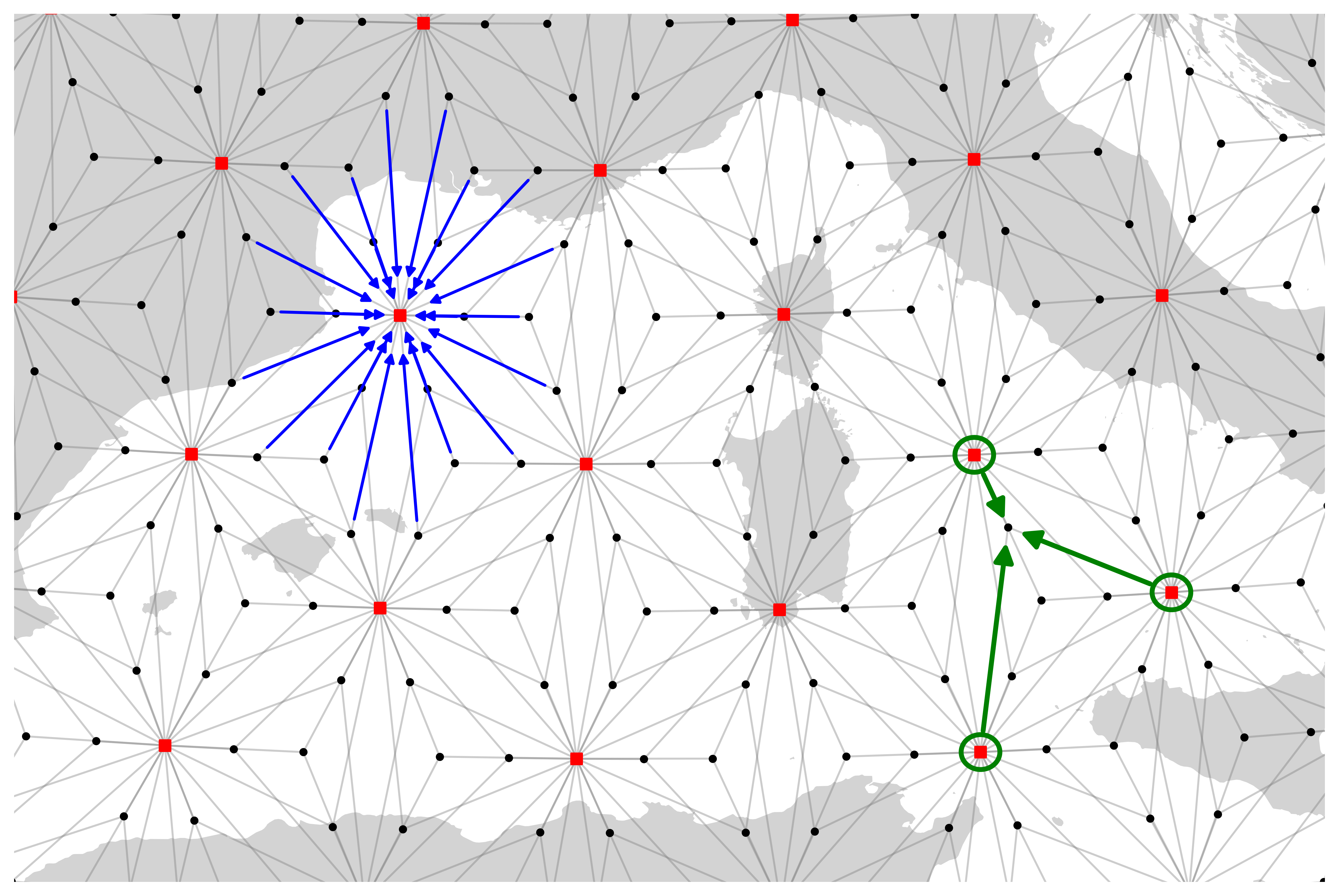}
    \caption{Examples of local graph structure (R03B04). Blue: a hidden node (red) connected to 24 data nodes (black dots) in the encoder. Green: Data node connected to its surrounding three hidden nodes in the decoder.}
    \label{fig:node_connectivity_map}
\end{figure}

A notable feature of the AICON model (apart from its training dataset) is its graph construction directly based on ICON’s triangular meshes. The GNNs operate on ICON's so-called $RnBk$ grids, which originate from an icosahedron whose edges are initially divided into $n$ parts, followed by $k$ subsequent edge bisections~\cite{prill2025}. This hierarchical grid construction process creates two additional fields that keep track of the refinement level of every grid vertex and cell.
The AICON model makes use of these evenly distributed point sets of varying density as follows: Firstly, the hidden mesh of the processor is constructed as the union of these sub graphs. It therefore mixes long-range and short-range edges similar to the GraphCast multi-mesh~\cite{lam2023learning} and the number of hidden mesh nodes corresponds to the number of vertices in an $RnBk$ ICON grid. The set of edges in the hidden mesh is equivalent to the union of all edges in the coarse-to-fine level hierarchy $RnB0, \ldots, RnBk$. The hidden mesh (R03B03) is visualized in figure \ref{fig:hidden_mesh_aicon} where the hierarchy of refined meshes becomes apparent.   

Furthermore, both the bipartite encoder graph and the decoder graph are defined by the ICON grid topology (the decoder graph is defined as the transpose of the encoder mapping):  Every data node is connected to the hidden mesh vertices that are given by the cell-to-vertex relationship. Therefore every data point is connected to exactly three hidden mesh nodes, whereas each hidden nodes connects $24$ data nodes (see figure \ref{fig:node_connectivity_map}).
The details of the graph topology for all three refinement levels that are used are summarized in table \ref{tab:aicon_graph_specs}.

\paragraph{Model architecture.}

The actual AICON model instance is built on this ICON grid topology and consists of graph transformer networks that apply transformer-style attention to graph-structured data~\cite{shi2021} which are part if the Anemoi code library. 
It consists of graph-attention layer in the encoder, processor and decoder components. The processor
comprises 16 attention layers, the encoder and decoder feature only one each. Each layer uses multi-head attention with $16$
attention heads. 
Furthermore chunking is applied to the encoder and decoder to limit concurrent memory consumption, using two chunks currently.
The encoder prepends an embedding step where the feature dimension of the nodes and edges is lifted to the hidden dimension
of $512$. The nodes have $2\times 93$ input features (the meteorological variables from two time steps). The decoder
then in the last step projects down to the output dimension of $79$. 
All edges use edge length and direction as features and are also embedded to the hidden dimension. \\

Note that AICON/Anemoi uses so called additional trainable features which exist as either node-based or edge-based fields. These can be thought of as learnable, constant forcings or geospatial information and are simply included as additional features in the node and edge feature vectors. 
In particular $n_\text{trainable} := 8$ trainable features are used for each input node, hidden node (processor and decoder each) and every edge. 
For AICON's  $RnBk$ grids this amounts to

\begin{align*}
n_\text{trainable,total}
= 20 \;n_\text{trainable}\; n^{2}\,\Bigl(7\cdot 4^{k} \;+\; 5\cdot 4^{k-1} \;-\; 1\Bigr)
\end{align*}
additional trainable features.

The most important architectural parameters of AICON are summarized in table \ref{tab:aicon_model_specs}. It is worth noting that about $75\%$ of the trainable parameters are additional trainable features. \\

With the definition of the numbers of input and hidden nodes ($N_{input}$, $N_{hidden}$) as well as the input and hidden feature dimensions
($d_{input}$ and $d_{hidden}$) we can determine the compression factor 

\begin{align} 
    CF = \dfrac{N_{hidden} \times d_{hidden}}{N_{input} \times d_{input}} = \dfrac{368\,642 \times 512}{2.94912\times 10^6 \times 93} \approx 0.68.
\end{align}
So the compression is not very strong, indicating that the architecture focuses more on being expressive to describe fine details
than on strong abstraction.

\subsection{Anemoi software framework}\label{sec:anemoi-framework}

The Anemoi Framework is an open‑source, collaborative platform for data‑driven weather forecasting developed by ECMWF together with multiple Member State NMSs through a European collaborative initiative \cite{aifsblogAnemoi,Wijnands2025Anemoi}. Anemoi is specifically designed to support the development, training, evaluation and operational deployment of large‑scale machine learning models for meteorology. Its objectives span enhanced forecast accuracy, high spatial and temporal resolution, probabilistic forecasts with strong skill, and operational readiness achieved through scalability, efficiency and reproducibility.
Anemoi is written in Python and leverages popular machine learning and scientific packages including PyTorch \cite{pytorch2024},
PyG \cite{fey2025pyg20scalablelearning},
Numpy \cite{harris2020array},
Hydra \cite{Yadan2019Hydra},
Earthkit \cite{ecmwf_introducing_earthkit_2024},
Zarr \cite{alistair_miles_2020_3773450},
PyTorch Lightning \cite{Falcon_PyTorch_Lightning_2019},
mlflow \cite{Zaharia_Accelerating_the_Machine_2018},
Pydantic \cite{Colvin_Pydantic_Validation_2025}, and others.

Anemoi follows an object-oriented architecture that supports the instantiation of components from configuration files, providing a high degree of flexibility. Model architectures, data readers, preprocessing steps, training schedules, and inference pipelines can be specified in YAML-based configuration files, a widely used human-readable format for structured configuration data.
To exploit the above-mentioned inherent hierarchical structure of the ICON data in the model processor, we contributed new building blocks for the definition of graph nodes and edges to Anemoi which are now available as configuration options.

For efficient training, we converted the ICON-DREAM data for training into Anemoi-compatible Zarr datasets.
The physical fields, which will be described in Section~\ref{sec:training-data}, are stored in a 4D hypercube.
Its dimensions are time, feature, ensemble, and grid cell.
The feature dimension stacks all single-level and 3D variables.
The ensemble dimension serves as a technical placeholder for future extension to ensemble training and has a length of 1 for AICON.
The cell dimension indexes all horizontal grid points on the Earth’s surface using a locality-preserving ordering that approximates a space-filling curve.
The \texttt{data} is chunked along the first dimension, with one file per time step.
Additional fields in the dataset provide a time vector, latitude and longitude coordinates, as well as statistic vectors with per-feature global and temporal maximum, minimum, mean, and standard deviation.
The dataset is completed with metadata describing the data resolution, ordered lists of the feature names, and information on whether features are provided on single or multiple levels.
The metadata is used in the training to configure variable-specific loss weighting and to construct the input for the operational AICON inference.

\section{Training}\label{sec:training}

Here we detail the actual training procedure. We begin by describing the variables selected from the ICON-DREAM reanalysis (section \ref{sec:icon-dream})
including preprocessing steps and normalization in section \ref{sec:training-data}. This is followed
by the definition of the used loss function in section \ref{sec:loss_function}, the transfer learning strategy in 
section \ref{sec:transfer_learning} and concluded with the training curriculum laid out in section \ref{sec:training_curriculum}.

\subsection{Preparation of training data}\label{sec:training-data}

The underlying reanalysis data set ICON-DREAM has been introduced in section \ref{sec:icon-dream}. Here
we detail the selection of variables used for training, some preprocessing steps, describe additional
forcing fields as well as the normalization of different variables. A complete overview of all used
fields in shown in table \ref{tab:training_vaiables}.

\begin{figure}[htbp]
\centering
  % AICON encoder/decoder graph visualization
  % "Visualization of height levels of the AICON model, together with the currently used ICON model levels and AIFS pressure levels from ECMWF (Lang, S., Alexe, M., Chantry, M., Dramsch, J.S., Pinault, F., Raoult, B., Clare, M., Lessig, C., Maier‐Gerber, M., Magnusson, L., Bouallègue, Z.B., Nemesio, A.P., Dueben, P., Brown, A., Pappenberger, F., & Rabier, F. (2024). AIFS -- ECMWF's data-driven forecasting system.)"
  % 
  % created by: F. Prill, DWD
  % creation date: 2026-05-29
  % tools: Python (Jupyter notebook)
  %        plot_aicon_vertical_levels/aicon_vertical_levels.ipynb
\includegraphics[width=0.9\textwidth]{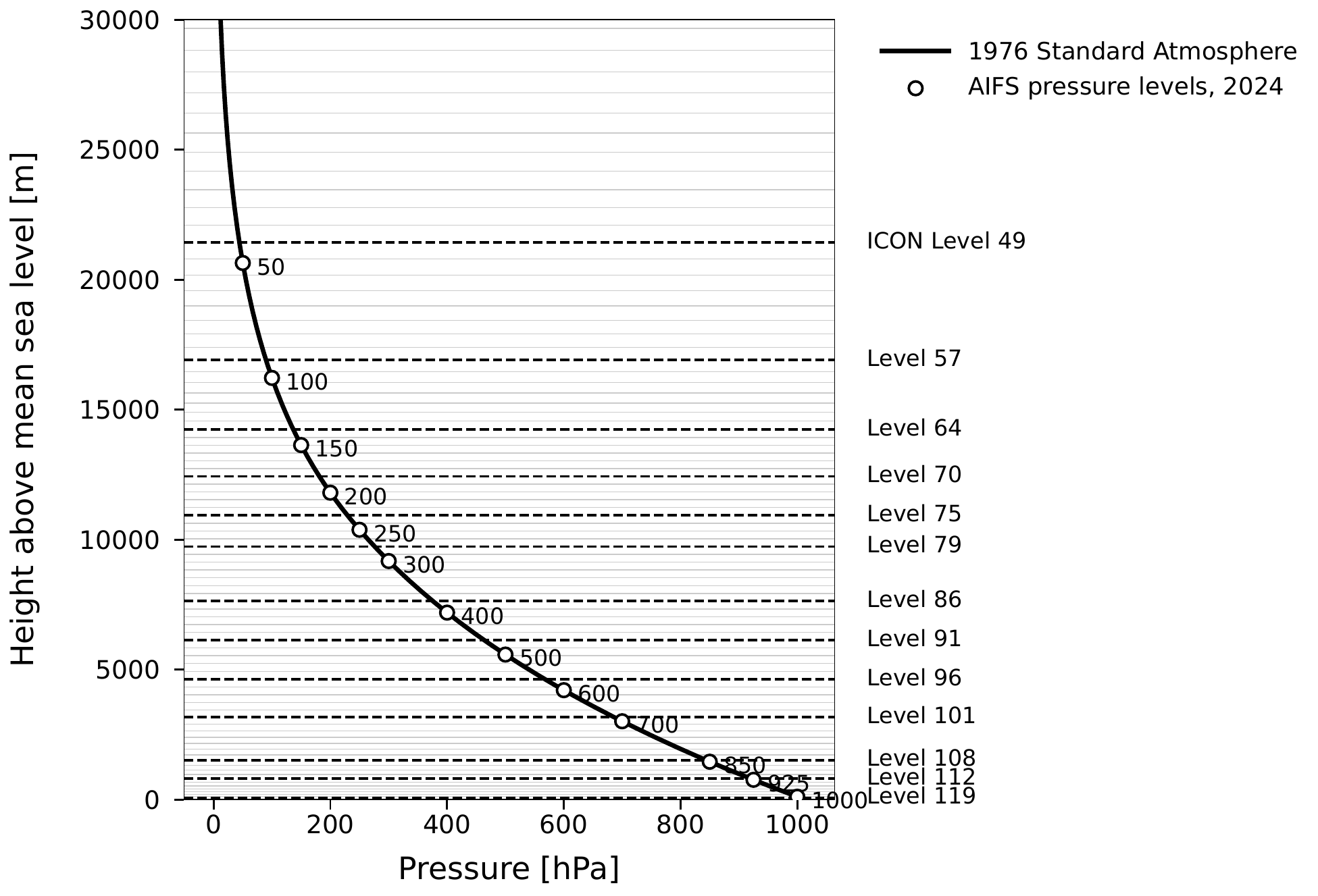}
\caption{Visualization of height levels of the AICON model, together with the currently used ICON model levels and AIFS pressure levels from ECMWF \cite{lang2024aifsecmwfsdatadriven}}
\label{fig:plot_aicon_levels}
\end{figure}

The most obvious dynamical parameters to train on are the prognostic variables defined
by the laws of atmospheric physics (see e.g. \cite{zaengl2015} for the prognostic
equations used in ICON), i.e. horizontal wind $U$ and $V$, temperature $T$, humidity $q_\text{v}$ 
and pressure $p$. From the corresponding 3D fields defined in the reanalysis process
(described in section \ref{sec:icon-dream}),
we take 13 levels (shown in figure \ref{fig:plot_aicon_levels}), in addition
to the corresponding (near-) surface components
($T_\text{2M}$, $U_\text{10M}$, $V_\text{10M}$, $p_\text{sfc}$, $RH_\text{2M}$) defined directly at the surface,
at two meter height or ten meter height. 
Further variables that are known two have strong
impact near the surface are soil temperature $T_\text{SO}$, soil moisture index $SMI$,
surface humidity $q_\text{v,s}$, and surface temperature $T_\text{G}$, which is equal to sea surface
temperature over the ocean. In particular the soil variables are known 
to memorize the initial state for several days to weeks beyond the short to medium forecast range
\cite{stacke2016lifetime} and are thus important
predictors for future weather states. The SMI needs some special preparatory steps. The available variable
from the ICON model output (within the ICON-DREAM reanalysis, described in section \ref{sec:icon-dream}) is
soil water $W_{SO}$ in units of $kg/m^2$ for each of eight different soil levels separately.  Since the
impact of soil water on the atmosphere depends on the soil properties (which are discrete labels and thus possibly difficult to learn)
the SMI lends
itself as a soil type independent description. It is defined as (see \cite{viterbo1995improved} and equation (3) in \cite{hunt2009development})
\begin{align}
    SMI = \dfrac{\frac{W_{SO}}{\Delta z \rho_w}  - PWP}{FC-PWP},
\end{align}
where $PWP$ is the permanent wilting point and $FC$ the field capacity which are both functions of soil type in ICON.
The density of water is taken as $\rho_w = 1000~kg/m^3$. $W_\text{SO}$ is accumulated over
the soil layers in between $0-3\,cm$ and $9-81\,cm$, such that $SMI$ is effectively a two-layer variable in the training data.
It assumes values in between $-0.93\leq SMI \leq 2.15$. Water points are defined with $SMI=2$. \\

Since all the above-mentioned variables are dynamic in time,
and are needed at every time step in the auto-regressive forecast inference, we treat
them as prognostic in the training. Further, there are forcing parameters that are either
constant in time (surface height $H_\text{sfc}$, land fraction $f_\text{land}$, lake fraction
$f_\text{lake}$, surface roughness $Z_\text{0}$, sub-grid scale orographic variability
$\sigma_\text{SSO}$ and  surface emissivity $\epsilon_\text{sfc}$, sine and cosine of latitude
and longitude) or have an analytical time-dependence (sine and cosine of "day of year"
and "hour of day" and insolation, which is treated as cosine of the solar zenith angle).
The three-hourly accumulated precipitation $TP$ is the only diagnostic quantity of AICON, i.e. it is the only variable that is not used as (auto-regressive) model input.

Most variables are normalized to zero mean and unit standard deviation in an initial model layer. Variables that are naturally in the range
$[0,1]$ or $[-1,1]$ (see table \ref{tab:training_vaiables}) are not normalized.
The positively definite parameters $f_\text{land}$, $f_\text{lake}$, $Z_\text{0}$, and $\sigma_\text{SSO}$,
are rescaled to $[0,1]$. Precipitation is rescaled by its standard deviation
and further bound from below to be larger than zero by a ReLU activation in the final layer.

\subsection{Loss function}\label{sec:loss_function}

We use a weighted mean squared error loss (wMSE):
\begin{align}
    \mathcal{L} =\frac{1}{\mathcal{N}} \sum_{x,v,l} w_{x,v,l} (y_{x,v,l}-\hat{y}_{x,v,l})^2,
    \label{eqn:loss_function}
\end{align}
where $y$ is the model prediction, $\hat{y}$ is the target, $w$ are tunable weight factors and
$\mathcal{N}$ is a suitable normalization factor.
The sum runs over three indices: $x$ for space, $v$ for variable and $l$ for the vertical levels, in case a given 
variable has several. Here the spatial dependence is described by area weights, that just weigh a given node
by the fraction of the area of the globe it represents. The choice of variable specific loss weights in described in table \ref{tab:var_weights}. Further
there is the level dependent weighting applied for atmospheric 3D variables ($U$, $V$, $T$, $q_v$ and $p$) that puts stronger emphasize 
closer to the surface. The weight is is $1.0$ at the model top and increases to $5.0$ in the lowest model level (close to the surface). The values of
those weights were chosen to yield similar contributions to the loss function across all variables (each $(x,v,l)$) after the first training step.

\begin{table}[htbp]
    \caption{Variable weighting used in training, see loss function definition in equation \ref{eqn:loss_function}}\label{tab:var_weights}
    \centering
    \begin{tabular}{c|c}
        variable &  weight \\
        \hline 
        $U, V$  &   $0.1$\\
        $q_\text{v}$& $0.5$ \\
        $T$ & $1$ \\
        $p$ &   $15$ \\
        $p_\text{sfc}$ & 15 \\
        $U_\text{10M},V_\text{10M}$ & $0.1$ \\
        $RH_\text{2M}$ & $5.0$ \\
        
    \end{tabular}
\end{table}

If autoregressive multi-step rollout in training is applied the above loss function (equation \ref{eqn:loss_function}) is used for each step.
The model output $y_t$ of a given step $t$ is then fed into the model several times $y_t \rightarrow y_{t+1}$ and each $y_t$ is then used against the
respective target at the appropriate time. The resulting loss is simply the average of the losses at individual times 

\begin{align}
    \mathcal{L}_T([y_1,\ldots,y_T],[\hat{y_1},\ldots,\hat{y}_T]) = \dfrac{1}{T}\sum_t^T \mathcal{L}(y_t,\hat{y}_t),
\end{align}
where $T$ is the number of rollout steps in training.

\subsection{Transfer learning}\label{sec:transfer_learning}

\begin{figure}[htbp]
\centering
\includegraphics[width=0.95\textwidth]{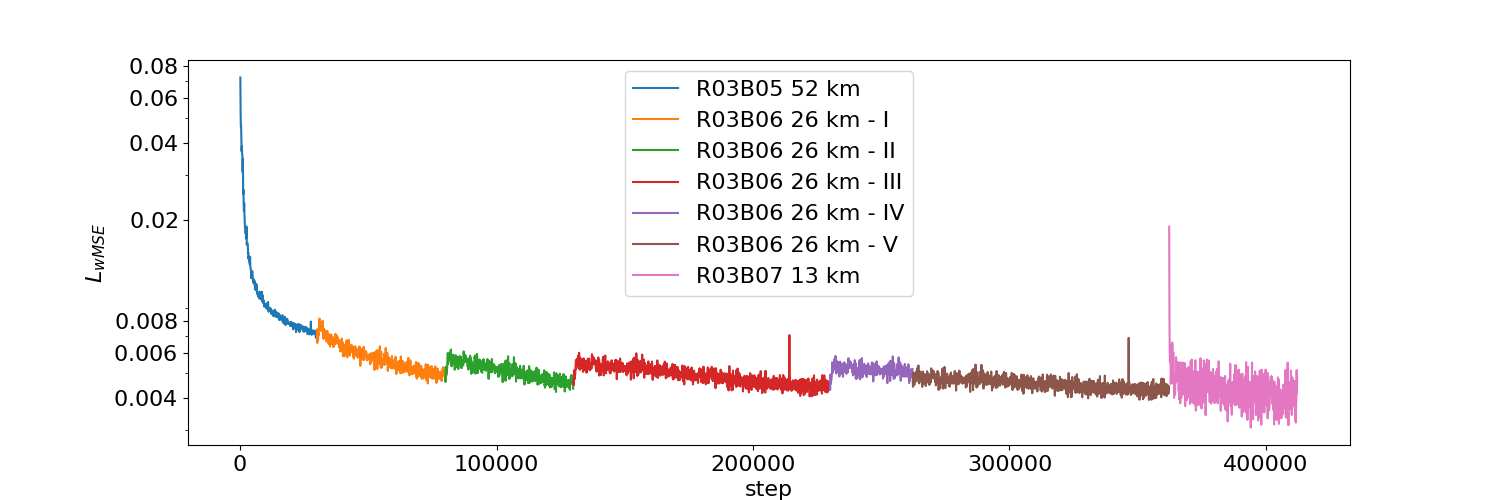}
\caption{Visualization of the training loss per step for all runs listed in Table~\ref{tab:training_runs}, arranged in a single row.}
\label{fig:plot_transfer_learning_train_loss}
\end{figure}

Here we shortly describe the details of our transfer learning protocol. Since the ICON-DREAM data set (section \ref{sec:icon-dream}) 
has a relatively high spatial resolution of 13~km and since AICON uses a hidden multi-mesh, the model should be able to generalize
to different resolutions. It thus seems a good idea to train the large scales first and the finer scales later to make the training computationally more
efficient. We thus create coarser training data sets by sub-sampling the 
full dataset. Each sub-sampling step reduces the average grid point distance by a factor $1/2$ and the number of input nodes by $1/4$. The number of hidden nodes
is always sub-sampled accordingly. We start our training from a data set that is sub-sampled twice, resulting in a number of input nodes that is $1/16$ th
compared to the full data set. The only thing that changes, as far as the model is concerned, is the number of nodes and edges in the graph (see Table~\ref{tab:aicon_graph_specs}).
The learnable model parameters that describe the various layers (see e.g. table \ref{tab:aicon_model_specs}) are left unchanged, but the trainable features have to be re-initialized
since they depend on the particular graph. Transfer learning has been successfully used within the Anemoi framework to pre-train a model on 
a coarser scale to improve data coverage in \cite{nipen2026regional}, where a high resolution regional model was pre-trained on global ERA5.
So these were completely different data sets and a different spatial and temporal coverage. Our case is different, we do transfer between
data sets that are simply spatial subsets of one another. So we expect a high degree of transferability. This is indeed what we find, as
can be seen in figure \ref{fig:plot_transfer_learning_train_loss}.

\begin{figure}[htbp]
\centering
\includegraphics[width=0.9\textwidth]{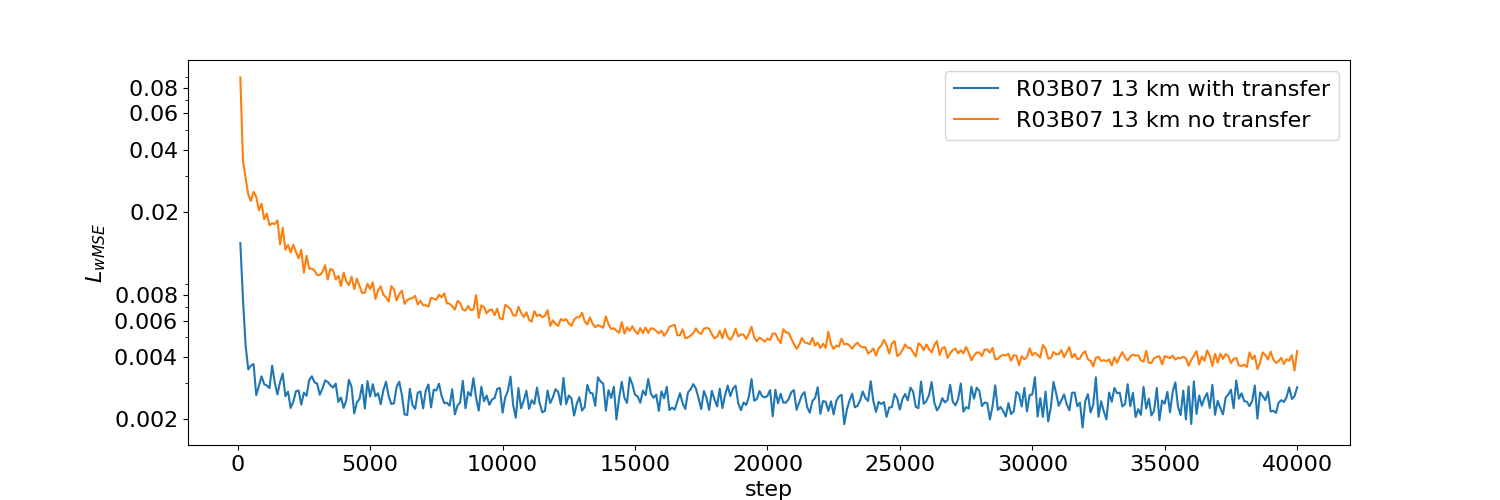}
\caption{Visualization of the training loss of a full resolution AICON (13 km) trained without transfer learning vs transfer learning by a restart from a 26 km checkpoint.}
\label{fig:plot_train_transfer_r3b7}
\end{figure}

Here the transfer from (R03B05 - 52~km) to (R03B06 - 26~km) is 
visualized by the first two graphs in that plot (blue to orange). The training loss is very well behaved across the transfer.
A direct comparison between a full resolution model (13 km) trained ab initio and a model fine-tuned via transfer learning from a 26 km checkpoint 
is shown in figure \ref{fig:plot_train_transfer_r3b7}. Both are trained with the same number of steps and it can be clearly seen that transfer
learning leads to a much faster convergence after $< 5000$ steps whereas the ab initio training has not converged to the same loss value even after
$40\,000$ steps.
In fact this training procedure saves a lot of time, since at the higher resolutions much less training is needed to converge. This training curriculum will be further
discussed in section \ref{sec:training_curriculum}

\subsection{Training curriculum}\label{sec:training_curriculum}

The training is executed on a node of eight NVIDIA A100-SXM4-80GB. When model sharding is applied, that is sufficient for even the full resolution data set.
For the two coarser data sets, sharding is not necessary.

As described in section \ref{sec:transfer_learning} a central concept of the training curriculum is the use consecutive training stages with progressively increasing spatial resolution.  In particular we start with a data set that has 52 km spatial resolution (R03B05 in the ICON nomenclature) and the transfer to R03B06 (26 km) and finally
the full R03B07 (13 km).  The training runs are summarized in table \ref{tab:training_runs} and the training loss is visualized in figure 
\ref{fig:plot_transfer_learning_train_loss}.

The higher resolution (26 km) allows for a smaller loss value toward convergence, than the coarser initial training run.
At 26 km resolution the model converges to a stable plateau, as indicated by the consecutive restart runs at resolution 26 km shown in . The plateau proved to be robust across five different restarts, during which the learning rate cycle was first
repeated, than elongated and finally the learning rate decreased. The loss always spikes at the beginning of the restarts to return to the value 
that was apparently limited from below by the resolution. This is proven by the last training stage, that implements the transfer to full resolution (R03B07 - 13 km).
Here the loss ($L_{13km}\approx 0.0041$) consistently drops below the value before ($L_{25km}\approx 0.0043$). This table also lists
a specific value of batch size for each training run. The coarsest run (R03B05 52~km) start with
a batch size of $32$ simply being the largest batch size possible given the hardware. 
The size $32$ is determined by using eight GPUs in training with four batches on each. For next resolution (R03B06 26~km)
only batch per GPU is technically possible such that only eight batches are used in total. And the full resolution (R03B07 13~km)
finally needs one model to be shared across all eight GPUs such in total only one single batch is used.

\begin{table}[ht]
  \caption{Training runs}\label{tab:training_runs}
  \begin{center}
    \begin{tabular}[c]{l|l|r|r|r|r|r}
    \hline 
    Resolution & name & epochs & steps & batch size &  hours run time & GPU h \\
    \hline
    R03B05 (52km) & \verb|rfl_5| & 23  & 30\,000 & 32 & 26.4 & 211  \\
    \hline 
    R03B06 (26km) & \verb|rfl_6_1| & 9   & 50\,000 & 8 & 44.5 & 356  \\
    R03B06 (26km) & \verb|rfl_6_2| & 9   & 50\,000 & 8 & 44.5 & 356  \\
    R03B06 (26km) & \verb|rfl_6_3| & 19   & 100\,000 & 8 & 91.7 & 733  \\
    R03B06 (26km) & \verb|rfl_6_4| & 6   & 32\,000 & 8 & 29 & 233  \\
    R03B06 (26km) & \verb|rfl_6_5| & 19   & 100\,000 & 8 & 91.7 & 733  \\
    \hline
    R03B07 (13km) & \verb|rfl_7| & 1   & 50\,000 & 1 & 34 & 269  \\
    \end{tabular}
  \end{center}
\end{table}

In addition to the training of the full resolution model just described the R03B05 (52km) checkpoint was also fine-tuned using multi-step rollout of 24 hours
(eight forecast regressions of three hours each) for $10\,000$ training steps. In this training mode the model is used auto-regressively and the loss function is the sum of all losses
for every single regression step. The influence of this fine-tuning will be investigated in section \ref{sec:rollout_vs_no_rollout}

\section{Operationalization}\label{sec:oper}

The transition of an NWP model, or in the present case an MLWP model, from scientific development to operational use is a crucial step for a NMS. Operationalization requires more than demonstrating forecast skill in experimental settings. It involves transferring a model into a production environment in which long-term stability, reliable execution, well-defined interfaces, monitoring, and timely product delivery are essential requirements. In such an environment, the model becomes part of an operational forecasting system that must function robustly on a sub-daily basis and provide dependable information for downstream users and decision-making processes.

In this regard, AICON entered technical operations (i.e., 24/7 routine operation of the model) on 2 September 2025, initiating an intensive evaluation phase by DWD's forecast department (see section~\ref{subsec:forecasters}). On 2 March 2026, AICON was officially introduced into operational use \cite{dwd2026aicon}. AICON is now an operational forecasting system and its current forecast products are documented in DWD’s official AICON Database Description \cite{DWD_AICON2026}. 

\subsection{Scientific and Technical Development and Evaluation}

The scientific and technical development phase were the base of AICON’s operational transition. Key activities included adapting the model code from its Anemoi base (e.g. application to ICON grids), developing and integrating pre- and post-processing components, and implementing interfaces to DWD’s forecast-production system. In particular, the inference code was rewritten de novo to exclude unused features, yielding a streamlined and operationally robust implementation. All modifications followed strict software-engineering practices, with most changes subject to the “four-eyes principle”.
An objective verification, using observations and comparisons with ICON
was carried out (see section \ref{sec:verification_observations}) confirming that AICON meets operational meteorological quality standards and delivers added value for forecasting.

\subsection{Routine Forecast Production, and Operational Environment}

The software and hardware infrastructure had to meet the stringent demands of routine operational forecasting. This included deploying a stable software environment, ensuring hardware availability with
redundancy and provisioning sufficient computational reserve capacity to absorb delays, and implementing continuous
 production monitoring and adapting the dissemination chain. To ensure reproducibility and mitigate risks of system failures or non-reproducible results, the model and its full software stack — including dependencies, preprocessing, and post-processing components — were packaged in a container. This containerization eliminates platform-specific variability, supports consistent deployment across development, testing, and operational environments, and enables reliable version control that is critical for robust and auditable operational forecasting. Bit-wise reproducibility, which is standard in classical NWP, could only be achieved with very high computational costs by eliminating non-deterministic algorithms, changed random seeds and differences in multi-threading execution. But the benefits were not found to outweigh the computational costs, and bit-wise reproducibility was thus dropped for AICON.
The application of containers, introduced for the first time in DWD’s NWP tool chain through AICON, is now being considered for adoption in operational classical NWP models at DWD as well. 
Like for NWP models, the organization of continuous data streams, provision of backup systems, inclusion of standard verification framework interfaces, integration into operational databases, and downstream post-processing had to be prepared for operational usage.
Since all these components are under control of the operators 24/7, comprehensive documentation and troubleshooting procedures had to be prepared.

\subsection{Evaluation by Forecasters in Practice}
\label{subsec:forecasters}

In addition to the aforementioned objective verification, meteorological evaluation in daily forecasting practice was essential in assessing the practical value of AICON for operational forecasting and warning activities. Since September 2025, forecasters continuously assessed selected forecast fields and relevant weather events and compared AICON forecasts with operational global NWP forecasts from ICON as well as further NWP and MLWP systems. These comparisons helped to evaluate the added benefit of AICON in real forecasting situations and to assess its strengths and limitations relative to established forecast guidance. Such \textit{manual} evaluation is essential before full operational deployment at an NMS, which supports different user groups: critical infrastructure, which includes public safety (fire services, THW), aviation, energy, agriculture, and emergency response. Further details about this subjective verification can be found in the corresponding
evaluation section \ref{sec:subjective_evaluation}.

\subsection{Approval and Dissemination}

After the benefit of the AICON forecasts had been confirmed by objective and forecaster's manual verification for all stakeholders, AICON could become fully operational. As part of this step, the AICON forecasts were conducted in the full operational environment, after continuously running for several months near real-time in an operational environment under 24/7 supervision of the operators.

Internal and external customers depend on timely data, so a robust distribution framework had to be established. First, we created the data streams required to visualise AICON forecasts in NinJo~\citep{koppert2004ninjo}. Second, the forecasts were disseminated to a range of public services and released openly through DWD-OpenData~\citep{dwd2026aicon_opendata}, accompanied by a comprehensive technical documentation.

With the operational implementation of AICON, its forecasts became part of the basis for DWD's weather warnings, one of the agency's core responsibilities. DWD was among the first NMSs to make an ML-based forecast system fully operational.

\section{Evaluation}\label{sec:evaluation}

This evaluation section demonstrates AICON's forecast quality, focusing especially on the design characteristics that distinguish it from existing AI‑based models:
These are the high spatial and temporal resolution, the use of an ICON-based training dataset, the abstinence of rollout in training, and the involvement of the weather service forecasting section in the operationalization process.
This section encompasses the verification against observations in section \ref{sec:verification_observations}, the evaluation of the consistency of
subsequent forecast runs in section \ref{sec:verification_jumpiness}, a spectral analysis in section \ref{sec:spectra}, and a hurricane case in section \ref{sec:hurricane_tracks}.
It is concluded with a short account of the subjective evaluation of DWD forecasters in section \ref{sec:subjective_evaluation}. First, however, we conduct
an initial study about longer forecast horizons during training in section \ref{sec:rollout_vs_no_rollout} to inform the choice for the final setup.

\subsection{Initial study: effects of autoregressive multi-step rollout in training (52 km model)}\label{sec:rollout_vs_no_rollout}

Typical MLWP models use a long forecast horizon during training, either as a single step \cite{bi2022pangu,bodnar2025foundation} or as multi-step rollout \cite{lam2023learning,lang2024aifsecmwfsdatadriven}.  
In this section we examine an analogous setup for AICON.
To save computational resources, and as a proof of concept, this investigation is conducted on the coarser gridded R03B05 (52 km) AICON. To that end we use the model checkpoint
that was additionally fine-tuned on 24 hour rollout (eight three hour steps) during training, as described in section \ref{sec:training_curriculum}. Forecasts for a single month (June 2025) are conducted twice a day (00 and 12 UTC) for a forecast lead time of 180 hours.\\

\begin{figure}
    \centering
    \includegraphics[width=0.95\linewidth]{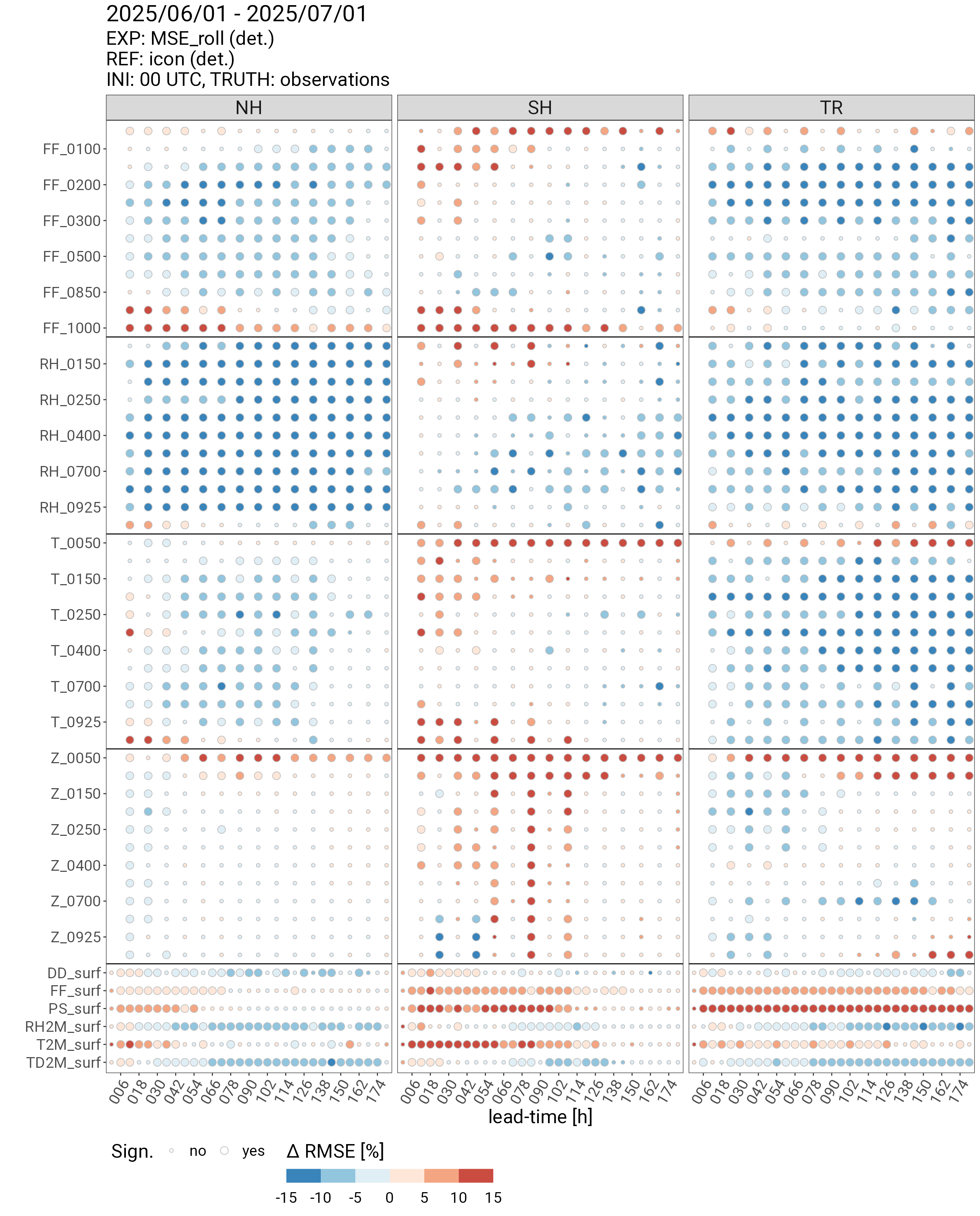}
    \caption{Relative difference in RMSE between AICON in R03B05 - 52 km configuration, with rollout training up to 24 hours - against operational ICON forecasts in June 2025 for the norther and southern extra tropics (NH and SH) and the tropics (TR). Forecasts are evaluated against synoptic stations at the surface and radiosondes (on different pressure levels). Blue: AICON performs better. Red: ICON performs better.}
    \label{fig:veri_r03b05_rollout_062025}
\end{figure}

We start with a verification against synoptic surface observations (SYNOP) and radiosonde observations (TEMP). 
The AICON (R03B05 52km - 24h rollout) forecasts are compared to the operational ICON forecasts as reference in Fig. \ref{fig:veri_r03b05_rollout_062025}. The score card shows the relative change in root
mean squared error (RMSE) conditional on forecast parameter and lead time for the Northern- (NH) and Southern Hemispheres (SH) and the tropics (TR). Blue colors indicate an improvement of AICON over ICON, red colors vice-versa. The size of the dots represents the outcome of a t-test for significant differences on a 95 significance level with larger dots used for significant differences. The AICON model (R03B05 52km - 24h rollout) has lower RMSE for most upper-air variables at most heights and most lead-times in the norther hemisphere and the tropics. 
The results for the Southern Hemisphere are ambiguous, showing both strengths and weaknesses, while the geopotential field generally yields neutral outcomes.
The near surface variables show lower performance at short lead times but become partially superior towards the end of the forecast range. The NH domain is most favorable for the AICON model. Surface pressure is the most difficult variable for AICON (see also section \ref{sec:verification_observations}). So clearly, by following the training strategy that has become standard in the mean time, we are also able to train a MLWP model that surpasses our operational physical model in the head line RMSE score for a large number of variables and lead times.\\

Longer training forecast horizons are known to cause a reduction of spatial small scale activity (see \cite{selz2025effective}). We thus present a spectral comparison of the current AICON (R03B05 52km - 24h rollout) with the AICON (R03B05 52km) it has been fine-tuned from and that has seen only the basic 3h time step in training in appendix \ref{sec:roll_out_spectra}. There we see that a reduction of the smaller scales is caused by the rollout fine-tuning. In particular figure \ref{fig:aicon_r03b05_rollout_ekin_ml6_spec_scales}
shows that this is the case for scales $< 1000\,km$. A detailed analysis of the spectral properties of the full resolution AICON model is presented in section \ref{sec:spectra}.\\

As one of our goals is to unlock the potential of the high resolution ICON-DREAM data set (see section \ref{sec:icon-dream}) as much as possible, based on the finding, that already a 52 km model shows clear signs of a reduction in effective resolution, we decided against using any sort of multi-step rollout.
In the next section we evaluate the full resolution AICON model that has been trained as summarized in table \ref{tab:training_runs}.

\subsection{Verification against observations}\label{sec:verification_observations}

The following verification is based on the version of AICON that became operational in September 2025 (see sections \ref{sec:model} to \ref{sec:training_curriculum}). A full year of AICON forecasts is available for evaluation, including an AICON re-forecast from March 2025 up to September 2025 and the operational forecasts to February 2026. For this evaluation we use only the 00  UTC runs with lead time of 180 hours. The forecast output is available in 3-hour increments up to 48 hours, and thereafter in 6-hour increments.  AICON forecasts are initialized with ICON analysis, therefore ICON forecasts of identical horizontal resolution serve as the verification reference.

\begin{figure}
    \centering
    \includegraphics[width=0.95\linewidth]{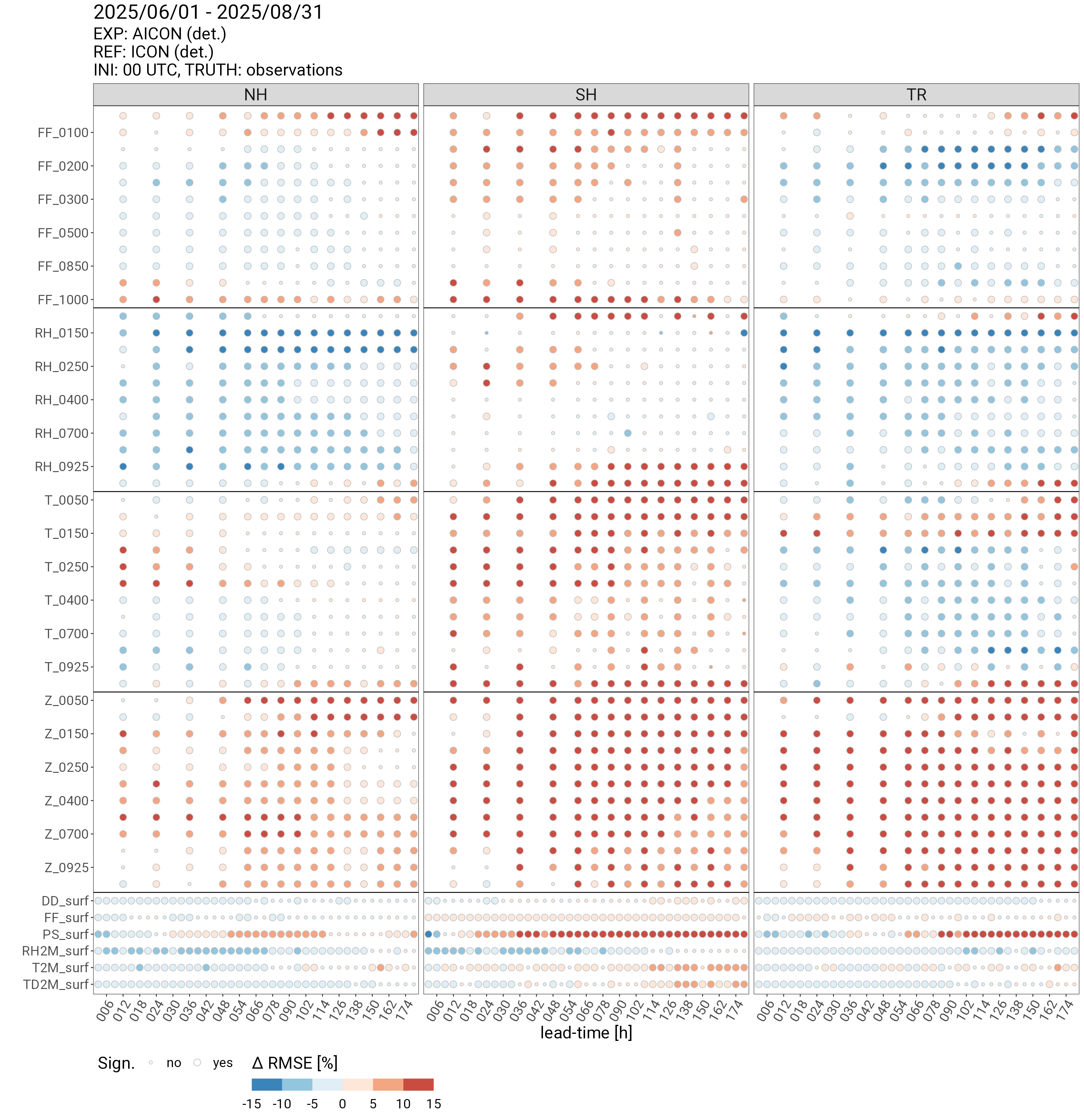}
    \caption{Relative difference in RMSE between AICON in it's operational configuration and ICON forecasts in June, July and August 2025 for the norther and southern extra tropics (NH and SH) and the tropics (TR). Forecasts are evaluated against synoptic stations at the surface and radiosondes. The size of the dots represents the outcome of a t-test for significant differences on a 95 significance level with larger dots used for significant differences.
    }
    \label{fig:veri_summary_jja}
\end{figure}

\begin{figure}
    \centering
    \includegraphics[width=0.95\linewidth]{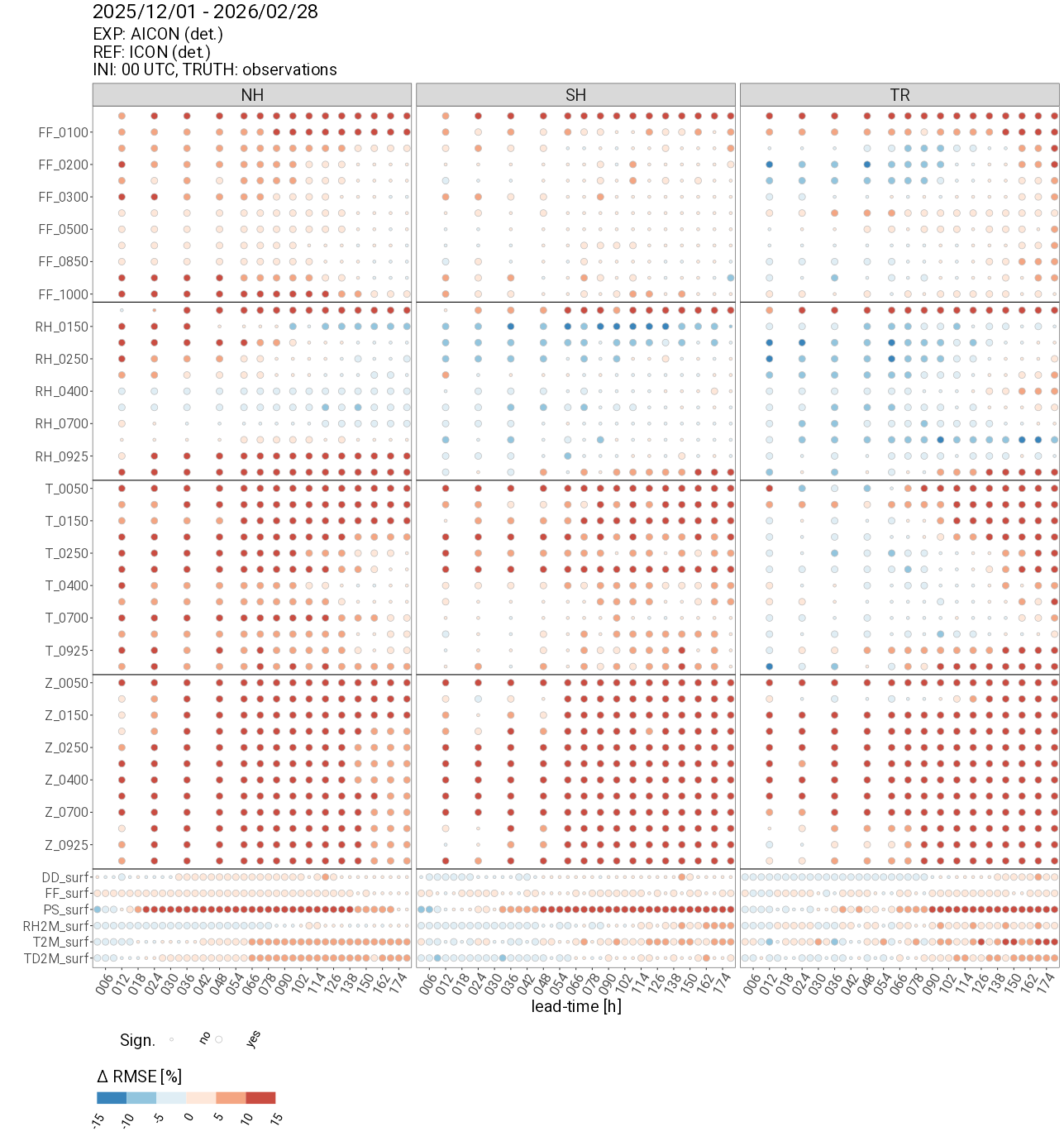}
    \caption{Same as Fig. \ref{fig:veri_summary_jja} but for December 2025, January and February 2026.}
    \label{fig:veri_summary_djf}
\end{figure}

To give a general impression of how AICON performs with respect to its NWP reference model ICON, score cards are shown for summer (JJA, Fig. \ref{fig:veri_summary_jja}) and winter (DJF, Fig. \ref{fig:veri_summary_djf}).  The verification is performed against SYNOP and TEMP observations as in the previous chapter. In terms of the RMSE the performance of AICON relative to ICON is mixed. On the positive side a gain in performance can be seen in most surface variables in the first forecast days. This gain is stronger and longer lasting in JJA than in DJF.  This is remarkable as the ICON forecasts especially near surface have shown to be exceptionally good (\cite{Zaengl_2023}). Also upper-air there is a clear performance difference between the seasons. AICON shows some quality improvement in forecasts of relative humidity (RH) and, in JJA also temperature (T) and wind speed (FF). On the negative side AICON seems to have difficulties in forecasting surface pressure (PS) at lead times beyond about day 2 and also the geopotential height (Z) scores significantly worse than ICON, especially in DJF. Also upper air T and FF seem to be more challenging for AICON in DJF. 
In addition to the relative scores, an investigation of the absolute score values is given in the Appendix \ref{sec:app_abs_scores}. Note that verification against observation requires vertical interpolation from model level to observation level. This interpolation error is part or the forecast error and stronger affects AICON with much less vertical levels. To circumvent this issue a verification against own analysis is provided in appendix \ref{sec:app_ana_ver} where the AICON improvement over ICON is significantly more pronounced.

\begin{figure}
    \centering
    \includegraphics[width=0.9\linewidth]{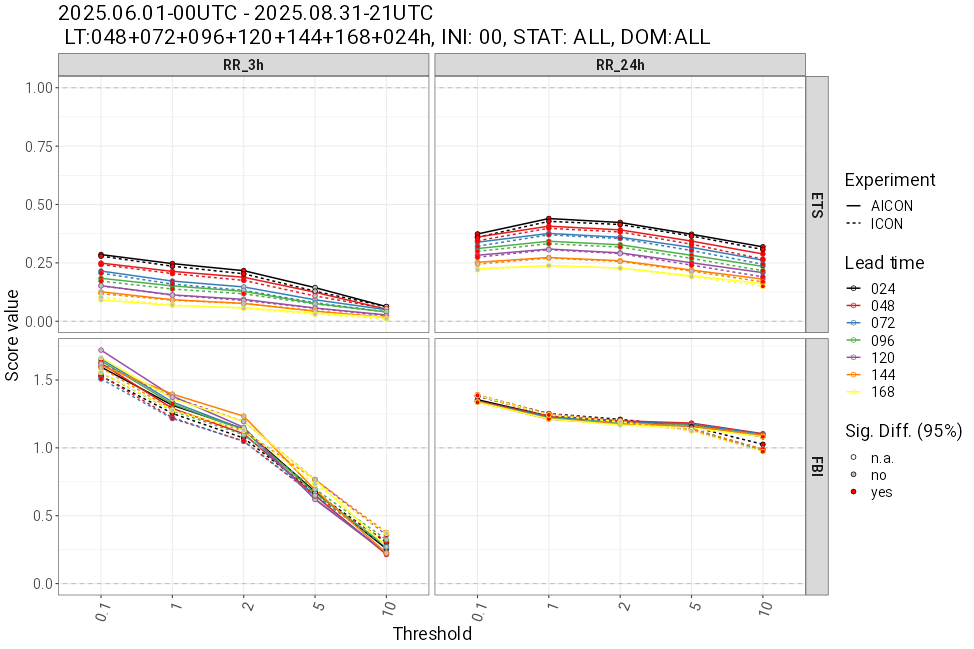}
    \caption{AICON and ICON categorical verification of 2025 JJA precipitation forecast. For 3 and 24 hourly accumulated precipitation (columns), different event thresholds (x-axis [mm]) and lead-times (colors), the ETS (upper row, higher is better) and FBI (lower row, optimum at 1) is shown for the global domain. }
    \label{fig:veri_precip_synop_categ_global}
\end{figure}

AICON provides total precipitation output which is evaluated categorically against SYNOP rain gauges. Figure \ref{fig:veri_precip_synop_categ_global} shows the AICON and ICON equitable thread score (ETS \cite{mason2011}) and frequency bias (FBI) for forecasting 3 and 24 hourly accumulated precipitation exceeding different event thresholds. Both models perform relatively equal. In terms of ETS significant differences can be mainly seen for the 24h precipitation sum for day 1 and day 2. Here AICON performs better for all thresholds meaning it is obviously more capable in discrimination rain from no-rain events. For 3 hourly precipitation and low event thresholds  as well as 24 hourly precipitation and high event thresholds AICON shows a significantly reduced frequency bias, bringing the FBI closer to the desired value of 1.

\begin{figure}
    \centering
    \includegraphics[width=1\linewidth]{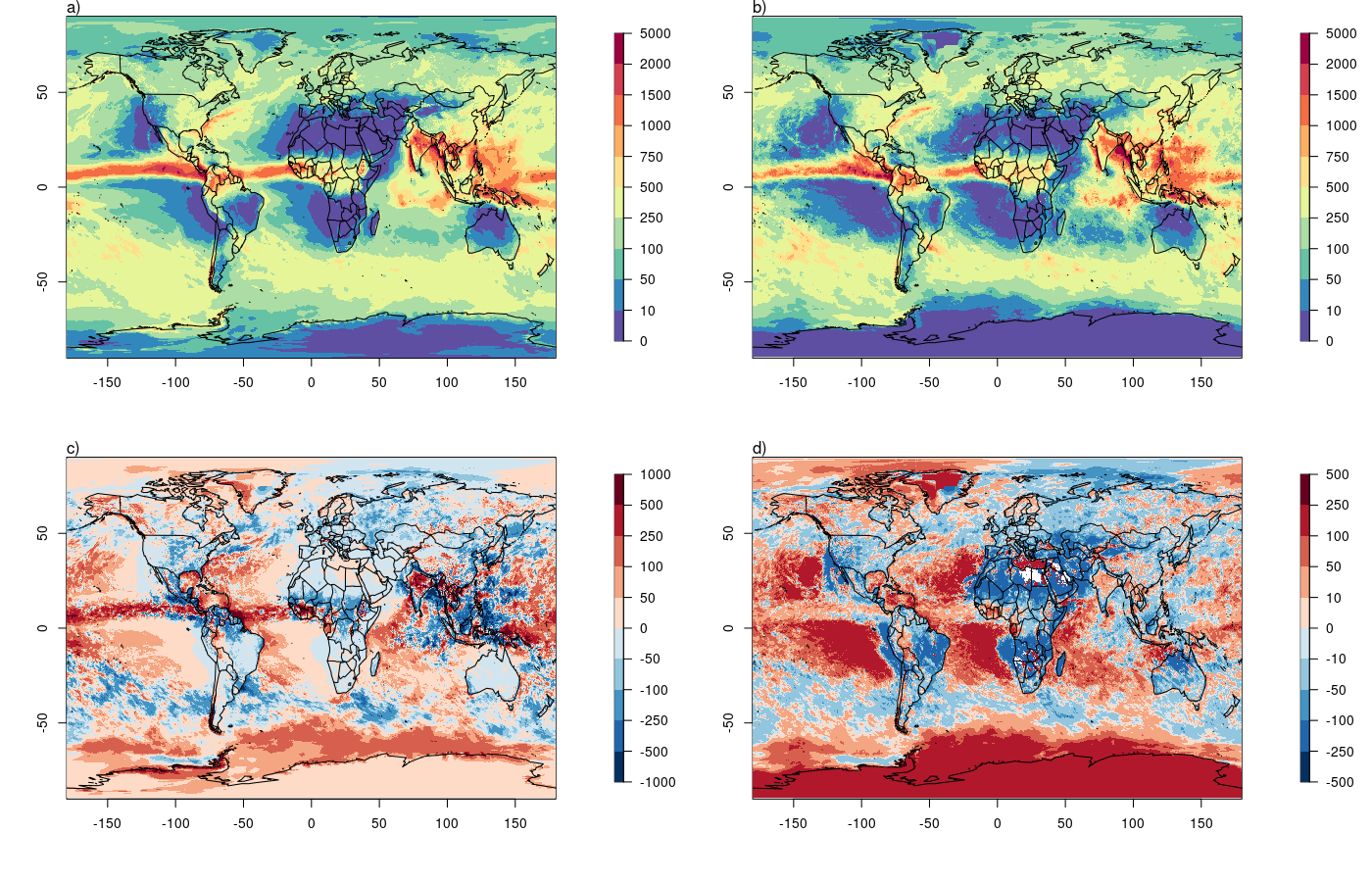}
    \caption{Accumulated total precipitation [mm] for JJA 2025 from AICON (a) and IMERG lev. 3 (b), differences [mm] (c) and relative differences [\%] (d) on a 0.25°x0.25° regular grid. AICON total precipitation is the sum of all 00 UTC,  0-24h forecasts valid in this period.}
    \label{fig:veri_prec_imerg}
\end{figure}

To give an visual impression of the quality of AICON precipitation forecast a comparison to IMERG level 3 data \cite{huffman2019} in the summer period (JJA 2025) is given in figure \ref{fig:veri_prec_imerg}. The top row shows the total accumulated precipitation on a 0.25° x 0.25° regular grid. At first sight AICON can reproduce regions of high precipitation as well as the arid regions in the sub-tropics and at higher latitudes. The lower row shows the absolute and relative differences of AICON to IMERG. The high precipitation amounts in the inner tropical convergence zone (ITCZ) over the oceans are mostly over-estimated by AICON by up to 50\%. Precipitation over land is mostly under estimated with exception in the arctic regions where AICON over-estimates the relatively small amounts of precipitation. Generally AICON precipitation appears more smooth then IMERG as can be seen by the lower degree of details e.g. in the ITCZ.  This behavior is also valid when accumulating AICON forecast of longer lead-times, however the spatial agreement is less accurate (not shown).

\begin{figure}
    \centering
    \includegraphics[width=0.6\linewidth]{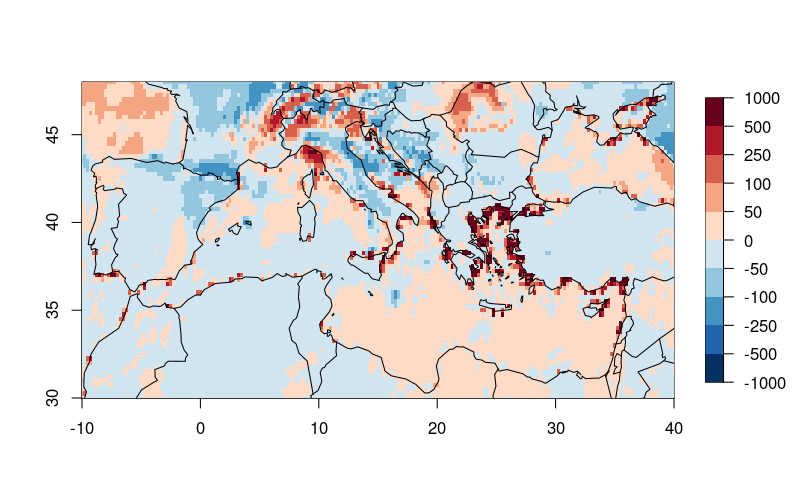}
    \caption{Total precipitation difference to IMERG [mm]  as in \ref{fig:veri_prec_imerg} c) but for 168h forecasts of AICON in the Mediterranean region. Small scale large positive differences to IMERG indicate the "rain-pox" problem, predominantly near coastlines, not present at shorter lead times.}
    \label{fig:veri_rain_pox}
\end{figure}

One undesirable property of AICON precipitation forecasts needs to be mentioned. At long lead-times, mainly along coast lines, precipitation forms in unrealistically high rates on the grid point scale. Figure \ref{fig:veri_rain_pox} shows this for the Mediterranean region. The small red grid boxes indicate a consistent occurrence of precipitation in AICON at always the same grid points. This happens only at coast lines however and it is not visible at shorter lead time forecasts of up to 72 hours. After about 72 hours the number of affected grid points as well as the amount of precipitation per time interval increase. 
Similar instabilities have been seen in early version of the  ECMWF AIFS, there this behavior could ultimately be changed by reducing the weight of soil moisture in training. For AICON investigations to alleviate these instabilities are ongoing, we cannot exclude that they affect precipitation only. 

\subsection{Jumpiness}\label{sec:verification_jumpiness}

\begin{figure}
    \centering
    \includegraphics[width=0.7\linewidth]{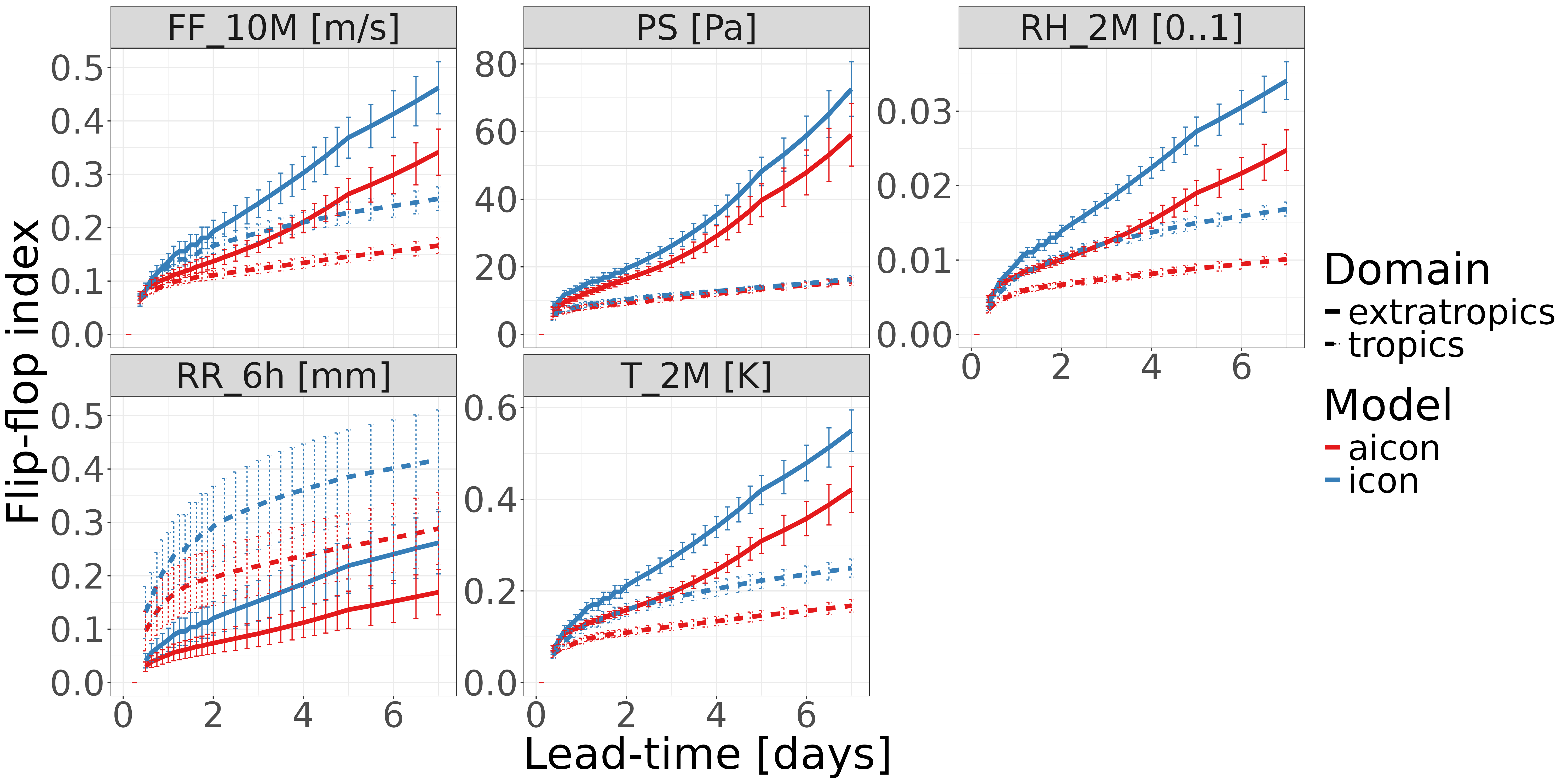}
    \caption{The flip-flop index for AICON (red) and ICON (blue) forecasts of surface variables. The index is calculated based on forecasts at surface observation sites in April 2026. Solid lines show the index for extra-tropical sites, dashed lines for tropical sites. The index standard deviation is shown by the error bars. Lower values indicate greater consistency between forecasts. }
    \label{fig:veri_flipflop}
\end{figure}

The term jumpiness is used to describe the alternating behavior of succeeding forecasts. When comparing forecasts with different lead time for a given valid time, the forecasts might change in a non-systematic way. The ideal would be a continuous convergence toward the observed truth as lead time approaches zero, but that is not always what is observed. A simple measure of this behavior is the flip-flop index, introduced by \cite{griffiths2019}. The flip-flop index basically measures the deviation of the forecast from a monotonic approach to the target that is the analysis or initial forecast step in units of the forecast variable. Forecast approaching the target monotonically are given a index of zero. Whenever a new forecast is further away from the target than the previous forecast, that difference between the succeeding forecasts is accumulated to the index. Here, the flip-flop index is calculated for AICON and ICON, for the surface variables wind-speed, relative humidity, surface pressure and precipitation, averaged over the northern extra tropics and the tropics (Fig. \ref{fig:veri_flipflop}). For all variables, domains and lead times the consistency between succeeding forecast is significantly greater for AICON than for ICON resulting in a lower flip-flop index. A forecast that is consistent from run to run can be experienced
as more trustworthy and reliable by the forecasters and users and be particular helpful in decision making.

\subsection{Spectral analysis}\label{sec:spectra}

As already discussed in section \ref{sec:rollout_vs_no_rollout}, MLWP models trained with an MSE-based loss function tend to exhibit lower forecast variability (also referred to as activity) compared to their physical counterparts \citep{benbouallegue2023risedatadrivenweatherforecasting}. Lower variability can lead to better scores in terms of MSE or RMSE at the cost of small scale fidelity. To investigate this further, a spectral analysis based on spherical harmonics is performed.

First, the data of ICON forecasts, AICON forecasts, ICON operational analysis (ICON ANA) and ICON-DREAM reanalysis on each model level separately have been horizontally interpolated from the original irregular R03B07 icosahedral grid to a full Gaussian F640 grid (with 1280 Gaussian-spaced latitudes and 2560 Gaussian-spaced longitudes, approx. 0.14°), using first order conservative mapping with cdo remapcon\footnote{See https://mpim-sw.gitlab-pages.dkrz.de/cdo/modules/man/Remapcon.html}. The resulting 2D global fields are transformed into spherical wave components after subtraction of the cell-area-weighted global mean value. The subtraction of the global mean ensures that the integral over the full spectra equals the variance. Power spectra are obtained by the sum of squared amplitudes of the components for each spatial scale to measure how much variance exists at different scales (wavelengths).

\begin{figure}
    \centering
    \includegraphics[width=0.49\linewidth]{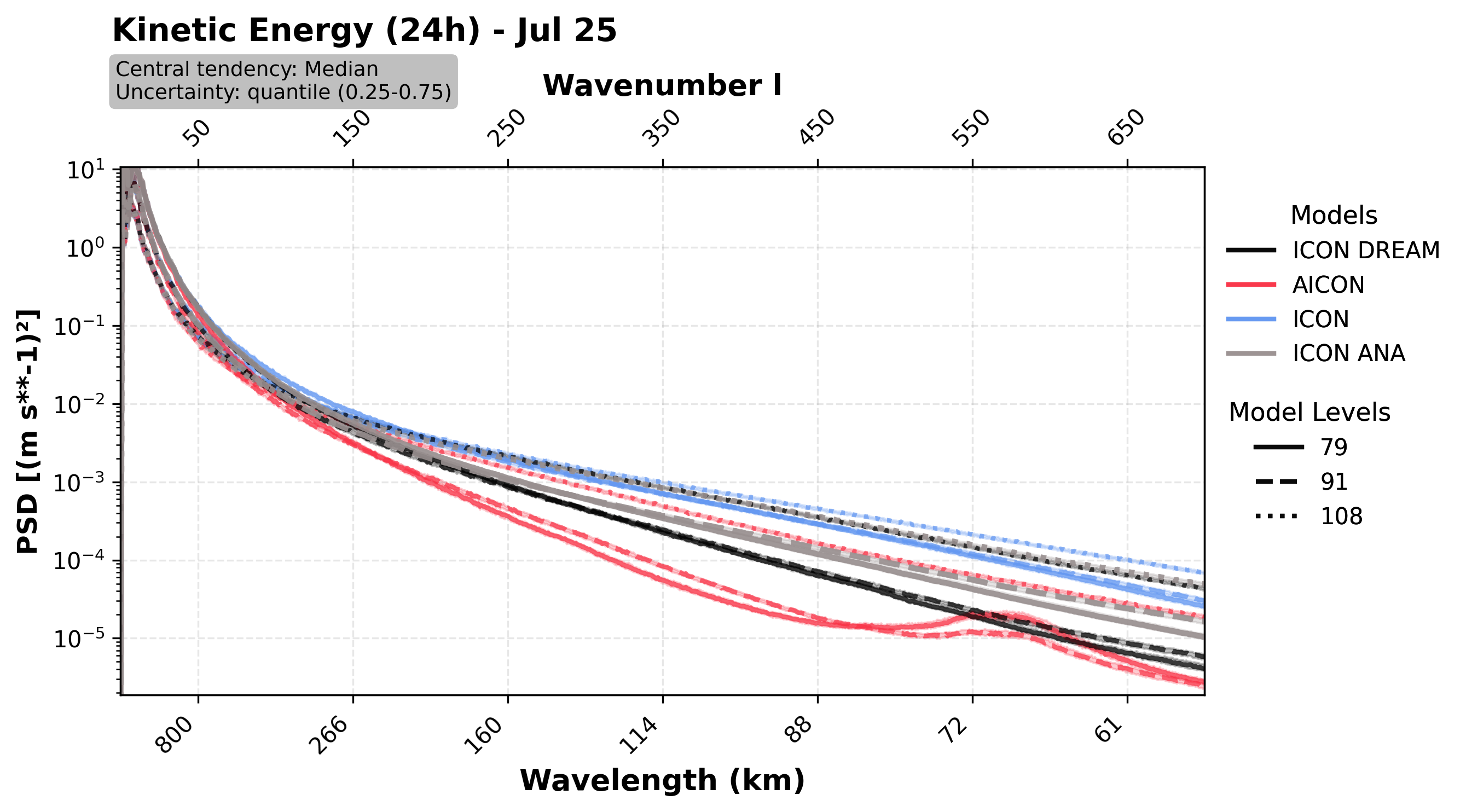}
    \includegraphics[width=0.49\linewidth]{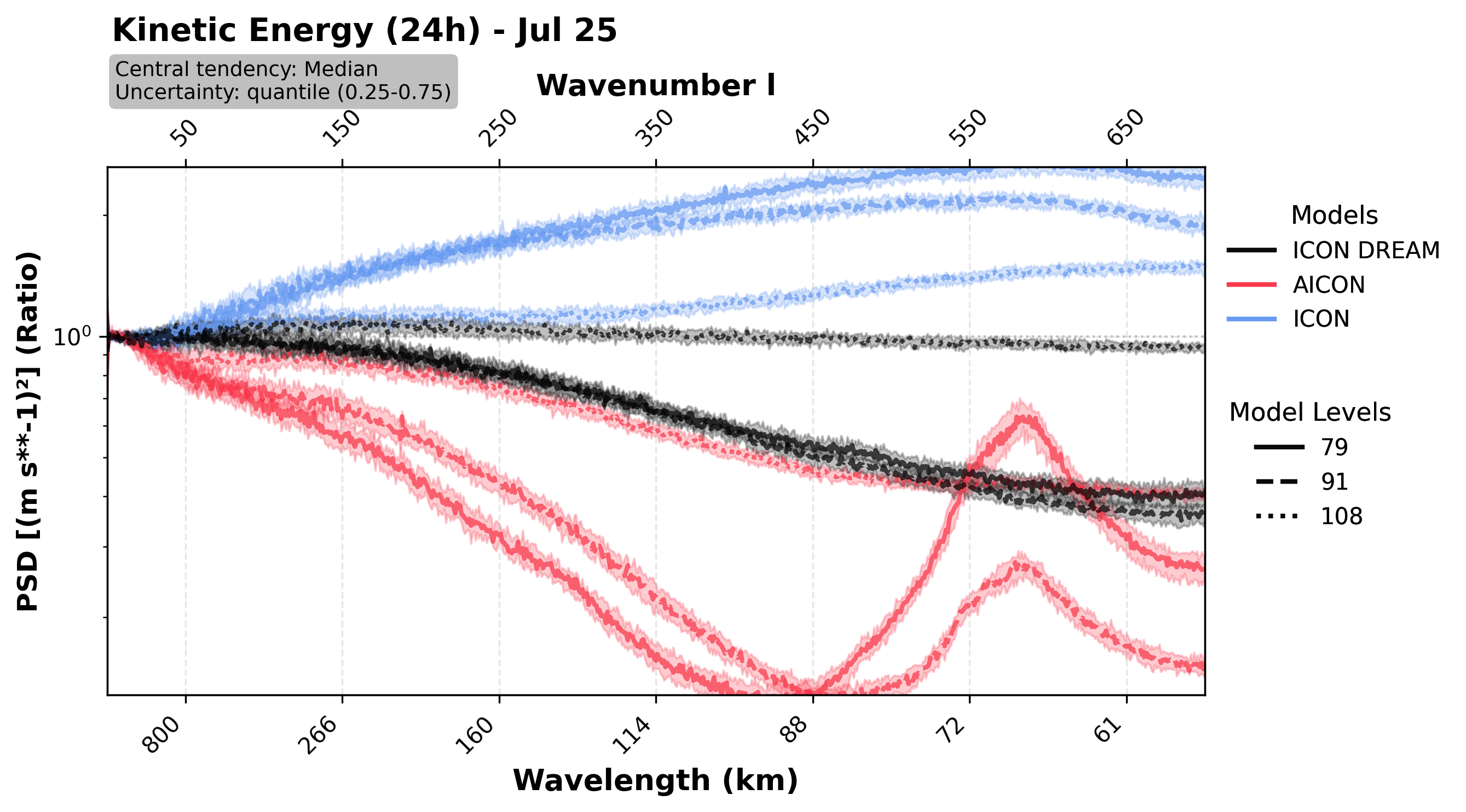}
    \caption{(left) Kinetic energy spectra on model levels 79, 91, 108 (corresponding to approx. 300, 500, 850 hPa) for forecasts from ICON and AICON, ICON analysis (ICON ANA) and the reanalysis ICON-DREAM. The lines represent median values over all 24 hour forecasts (initialized at 00:00 UTC) valid in July 2025. (right) Median of amplitude ratios with ICON analysis as reference.}
    \label{fig:power_spectra}
\end{figure}

Figure \ref{fig:power_spectra} shows kinetic energy spectra for model levels 79 ($\sim$ 300 hPa), 91 ($\sim$ 500 hPa) and 108 ($\sim$ 850 hPa), aggregated over all 24h-forecasts initialized at 00:00 UTC, valid in July 2025. Here the model level index refers to the ICON model level nomenclature. AICON has only 13 vertical model levels, but the ones corresponding to 
$[79,91,108]$ are included, see also figure \ref{fig:plot_aicon_levels}. 
ICON (blue) and AICON (red) show almost equal activity on larger scales > 250 km (left panel). Forecasts from the ICON model show power spectra similar to the operational analysis (ICON ANA), with even higher forecast activity on smaller scales. In contrast, AICON shows less forecast activity than ICON, ICON ANA and ICON-DREAM for all model levels. The gap in forecast activity is more pronounced for higher model levels ( $\sim 300$ hPa (solid) and $\sim 500 hPa$  (dashed)). Interestingly the direct comparison between the reanalysis ICON-DREAM and ICON ANA shows less activity in ICON-DREAM fields on smaller scales as well. This could partly contribute to the reduced energy spectrum of AICON, whose spectra seem to amplify the damping apparent in ICON-DREAM (right panel of figure \ref{fig:power_spectra} at least up to wave number $\sim 450$. The difference in spectral activity realism across vertical levels might be due to the influence of orography.
Orography, being constant in time, induces small scale predictability but its influence reduces with height. 
Another influence might come from  the level weights assigning more importance to lower levels during training. Thus errors on higher levels (potentially induced by noise) are not getting penalized as much as on levels closer to the surface.

Despite the lack of activity on smaller scales compared to ICON and own analysis data, the high resolution training data and training without rollout with 3h forecast steps results in a higher effective resolution of AICON fields compared to other state of the art AI weather prediction models (compare Figure 6 in \cite{mctaggart2026wp}). Hence AICON produces less blurry fields then other deterministic AI models trained mainly on ERA5 data using an MSE based loss function.

\begin{figure}
    \centering
    \includegraphics[width=1.0\textwidth]{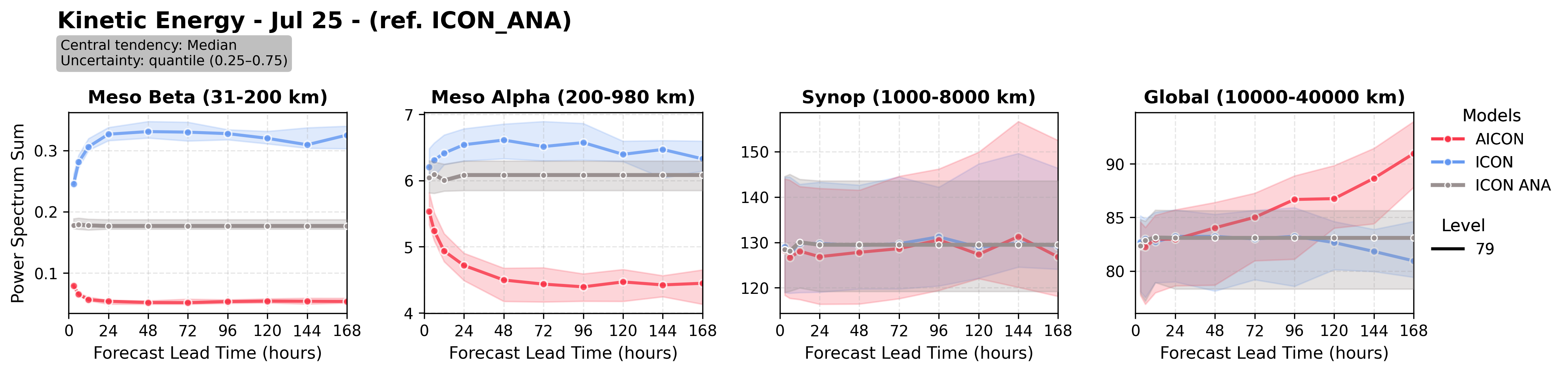}
    \includegraphics[width=1.0\textwidth]{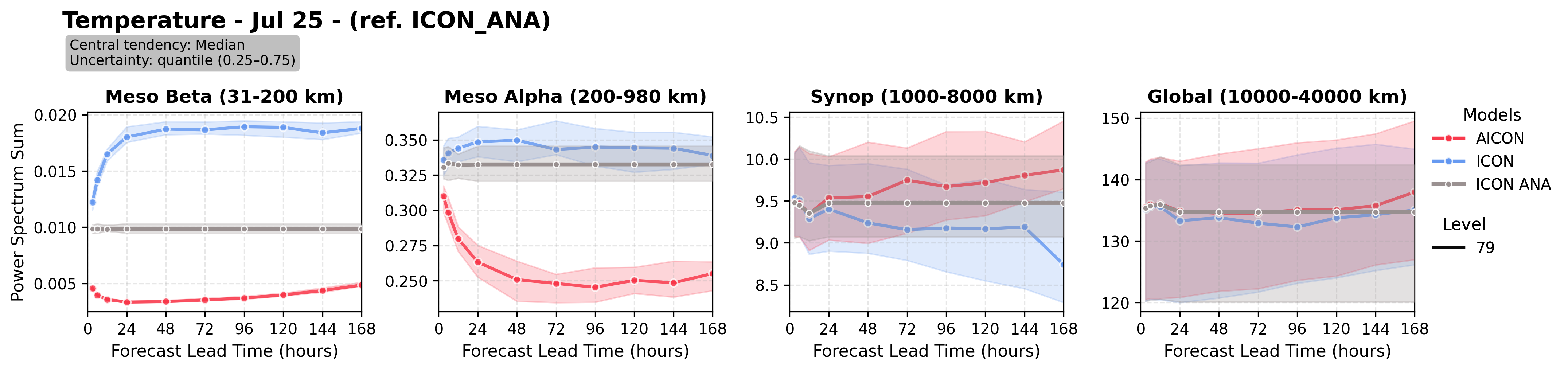}
    \includegraphics[width=1.0\textwidth]{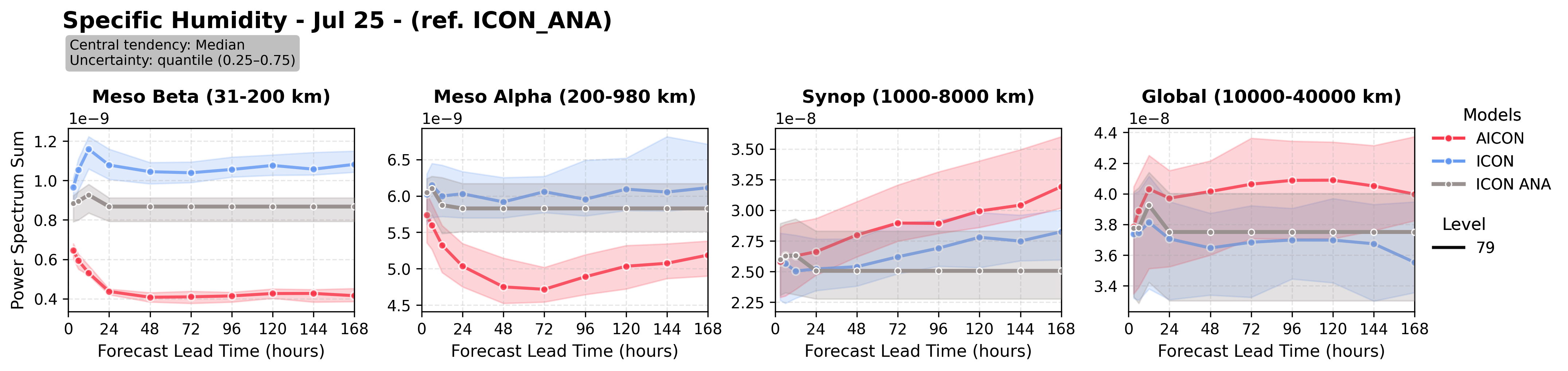}
    \includegraphics[width=1.0\textwidth]{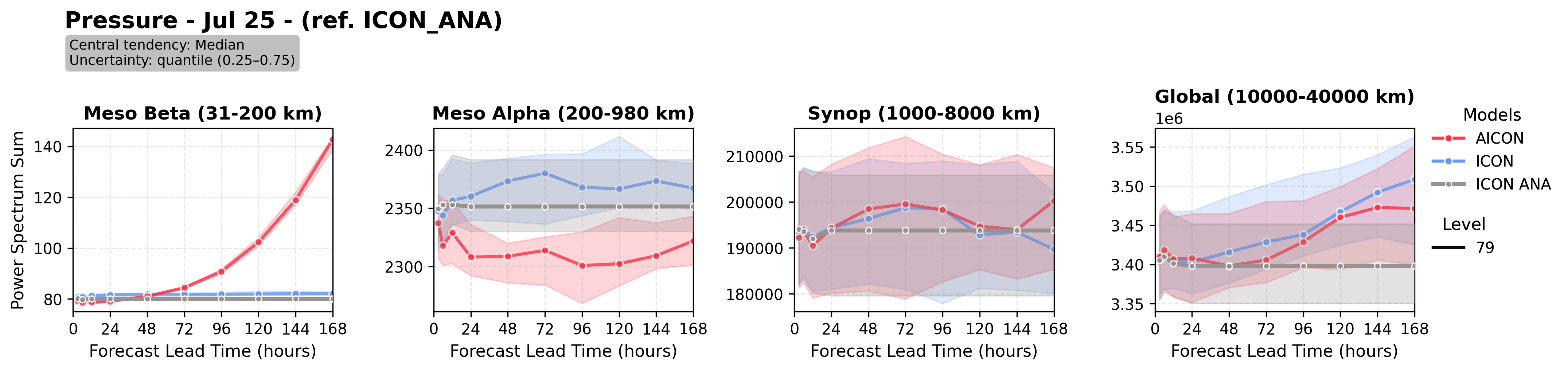}
    \caption{Scale dependent forecast activity over lead time for ICON and AICON for kinetic energy, temperature, specific humidity and pressure forecasts (model level 79, $\sim$ 300 hPa) issued for July 2025.}
    \label{fig:power_spectra_scale_lead}
\end{figure}

The scale dependent forecast activity over lead time is shown in figure \ref{fig:power_spectra_scale_lead} for kinetic energy (Ekin), temperature (T),  specific humidity (QV) and pressure (P) at model level 79 ($\sim$ 300 hPa) for ICON and AICON forecasts with ICON analysis as reference line (grey) valid in July 2025. The scale dependent activity is derived by aggregating the spectral components over different wave number intervals given in table \ref{tab:appendix_spectral_intervals}. On the mesoscales, the forecast activity of Ekin, T an QV increases for ICON during the first 24 hours of forecast lead time, while it is decreasing for AICON. 
Furthermore it can be noted that the day-to-day variability of spatial activity in AICON variable fields (Ekin, T, QV) represented by the inter quartile ranges (shaded colored areas) on meso-beta scales is reduced as well. This underlines the assumption of limited predictive skill on these scales. 
On the other hand on larger (synoptic and global) scales the activity of AICON fields is close to that of ICON for all variables shown here. 
The rapidly increasing activity on meso-beta scales in the pressure field (lowermost row) is related to the amplification of noise over lead time.
Its analysis and correction will be addressed in future work. This amplification of noise is potentially related to the relatively poor performance of AICON in 
pressure and geopotential, as already seen in the verification against observation in figures \ref{fig:veri_summary_jja} and \ref{fig:veri_summary_djf}.\\

It is worth to mention that the scale dependent investigation delivers new insights compared to an aggregation over all spatial scales. The behavior of forecast activity aggregated over all scales is dominated by the results on synoptic and global scales. However, as can be seen from figure \ref{fig:power_spectra_scale_lead}, the largest differences are obtained on mesoscales, indicating that the ML model learns the patterns on the larger scales very well, but lacks variability on smaller scales consistent with smoother forecast fields.

\begin{table}
	\caption{Spectral intervals of meteorological scales}
	\centering
	\begin{tabular}{lll}
		\toprule
		scale range & wave number l & wavelength [km] \\
        \midrule
        global scales & [0,4]& [40000,10000]\\
        synoptic scales & [5,40] & [8000,1000]\\
        meso alpha & [41,200] & [980,200]\\
		meso beta  & [201,1279] & [31, 200] \\
		\bottomrule
	\end{tabular}
	\label{tab:appendix_spectral_intervals}
\end{table}

\vspace*{2cm}

\subsection{Hurricane tracks}\label{sec:hurricane_tracks}

\begin{figure}
    \centering
    \includegraphics[width=0.5\linewidth]{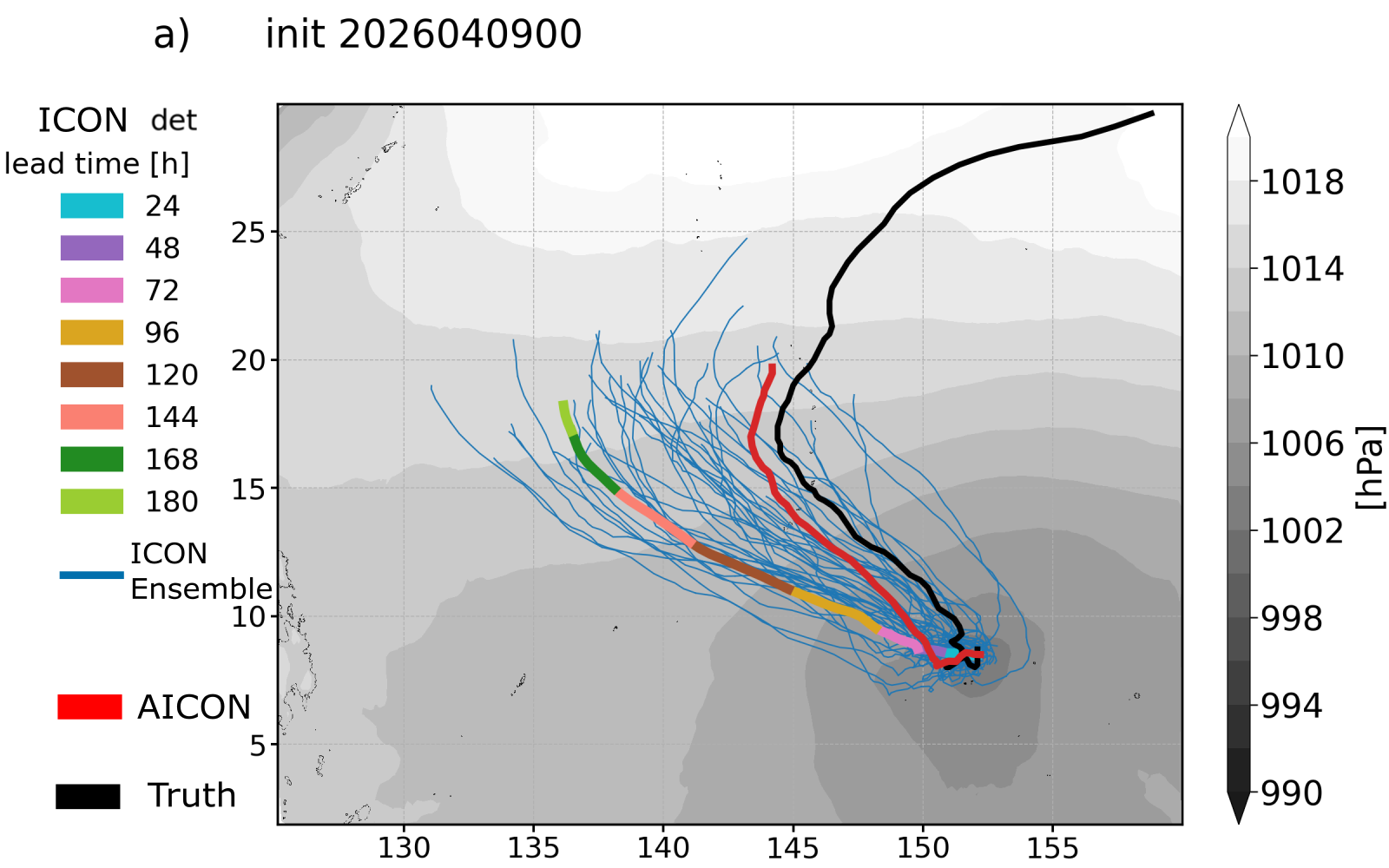}
    \includegraphics[width=0.45\linewidth]{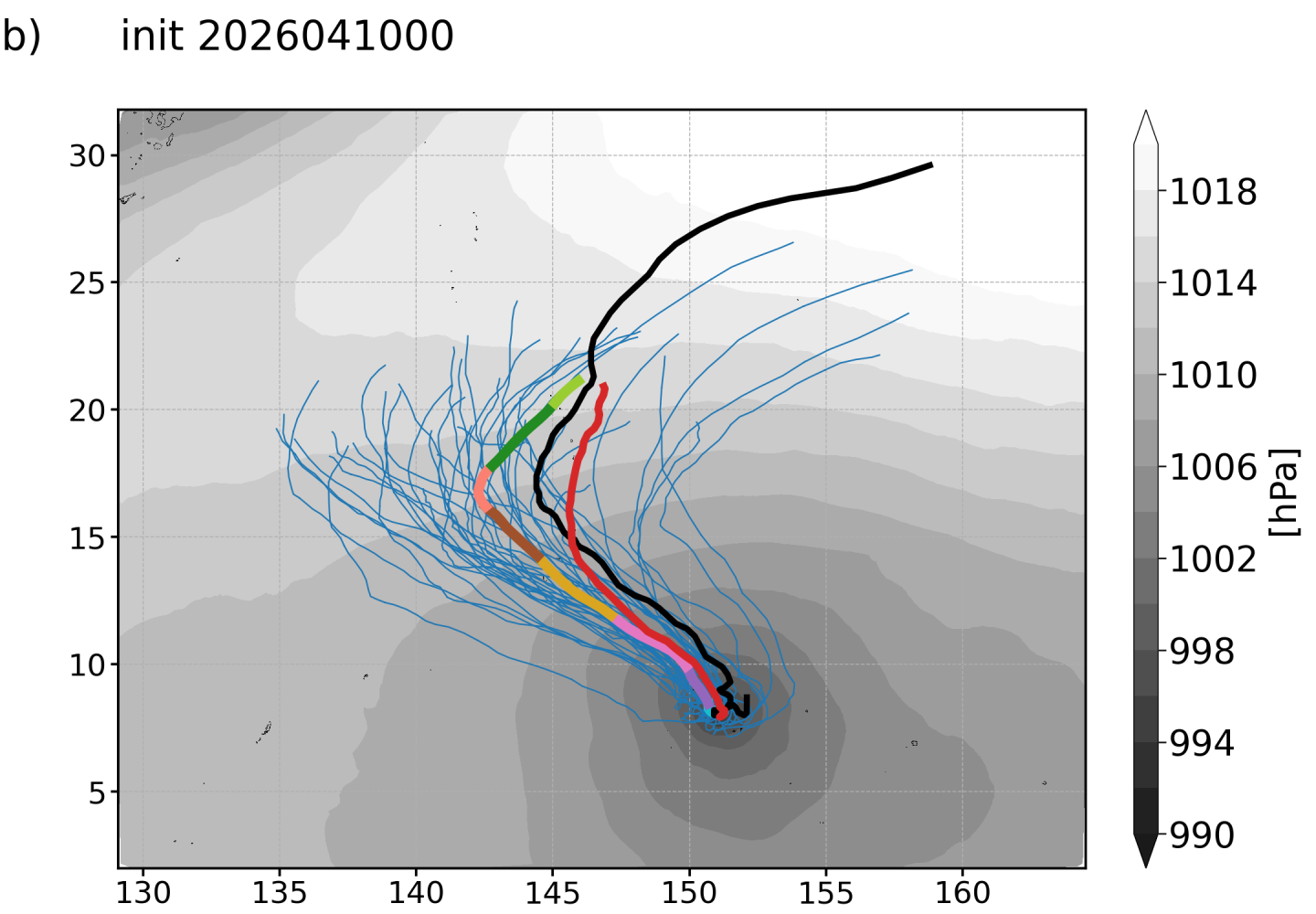}
    \hspace*{1cm}
    \includegraphics[width=0.45\linewidth]{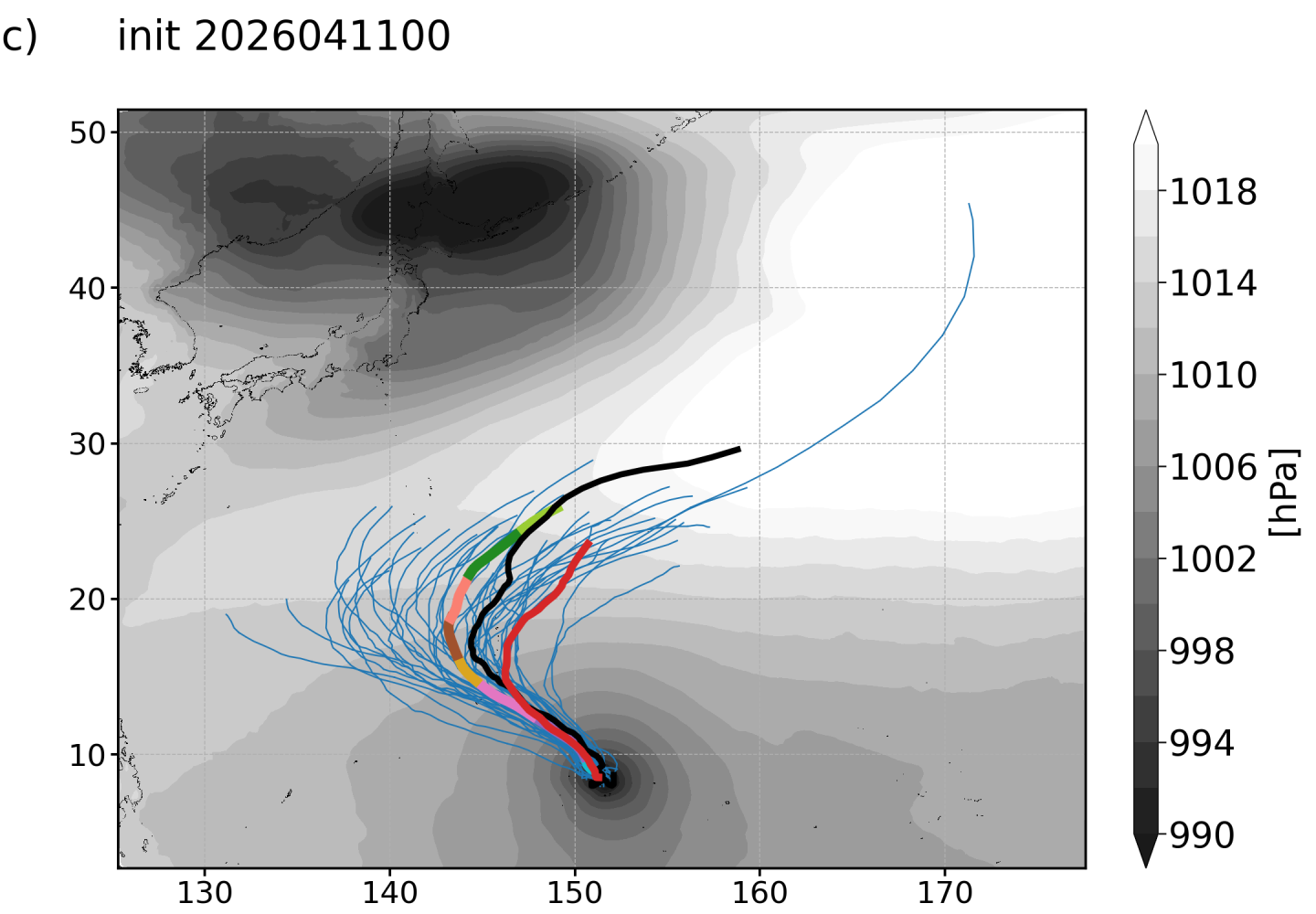}
    \includegraphics[width=0.45\linewidth]{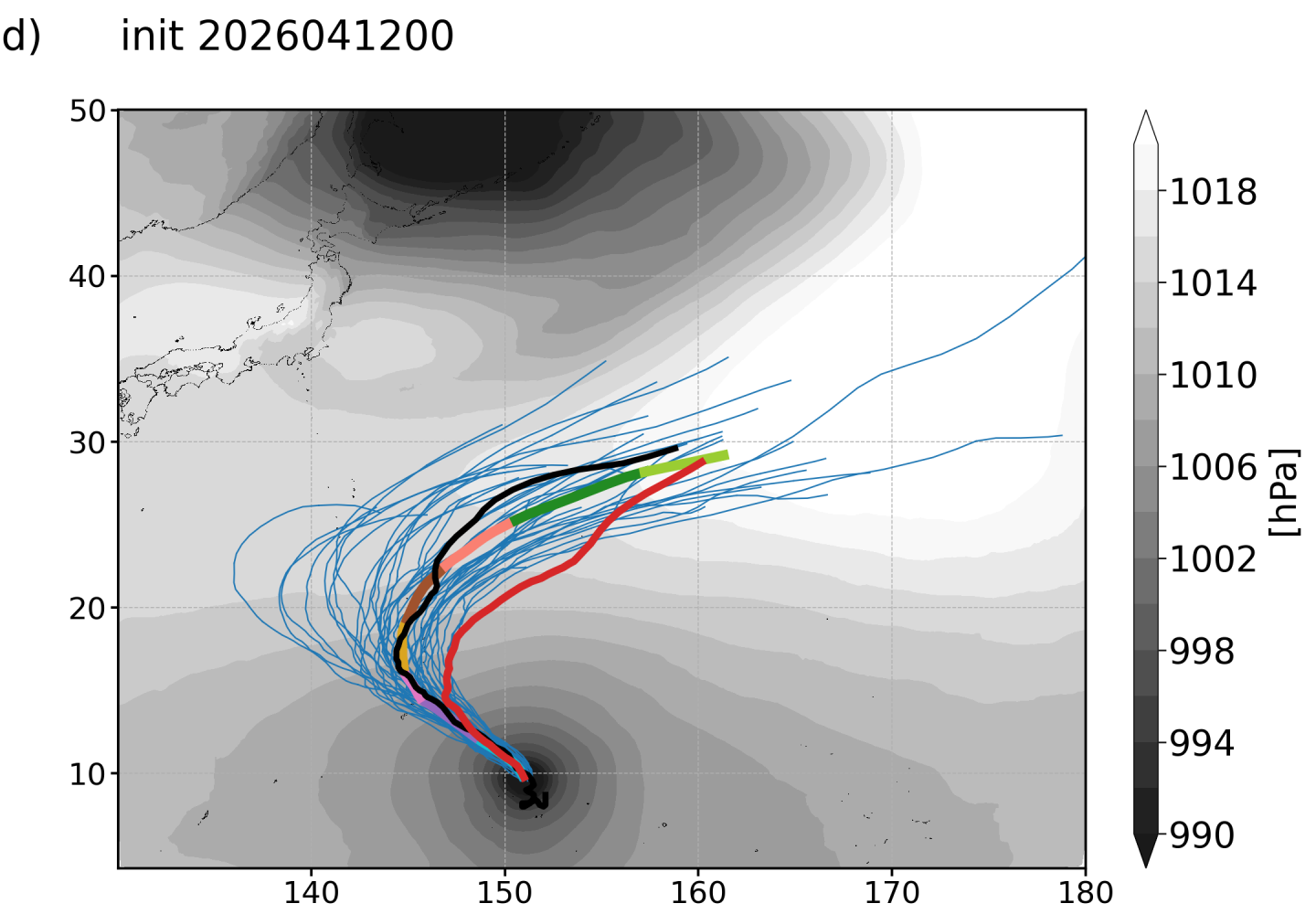}
    \caption{AICON tracks (red) of Typhoon Sinlaku (category 5) in comparison to the best track from the IBTrACS
    (International Best Track Archive for Climate Stewardship, \cite{knapp2010},\cite{gahtan2024}) data base (black) and cyclone tracks from the ICON deterministic NWP (time color). Member tracks of the ICON NWP Ensemble Prediction System (ICON-EPS) are blue. The different forecasts initialize on 9 (a), 10 (b), 11 (c) and 12 (d) of April 2026 at 00UTC, respectively. The legend on the upper left gives the colors used for the 24h lead time coloring of the ICON track. The background gray shades visualize the Mean Sea Level Pressure(MSLP) field of the initial ICON operational analysis. The domain settings of the different panels are given in degrees East/North. The panel titles show the $[$min MSLP, max wind speed (u,v), max gusts $]$ along the tracks for ICON and AICON. These numbers compare to the RSMC Tokyo - Typhoon center report \cite{JMA_RMSC} $905$ hPa as the minimum pressure and $215$ km/h for the 10-minute sustained wind and the JTWC report \cite{JTWC_USA} of 890 hPa and 285 km/h 1-minute sustained winds.}
    \label{fig:SINLAKU_tracks}
\end{figure}

\begin{figure}
    \centering
    \includegraphics[width=0.49\linewidth]{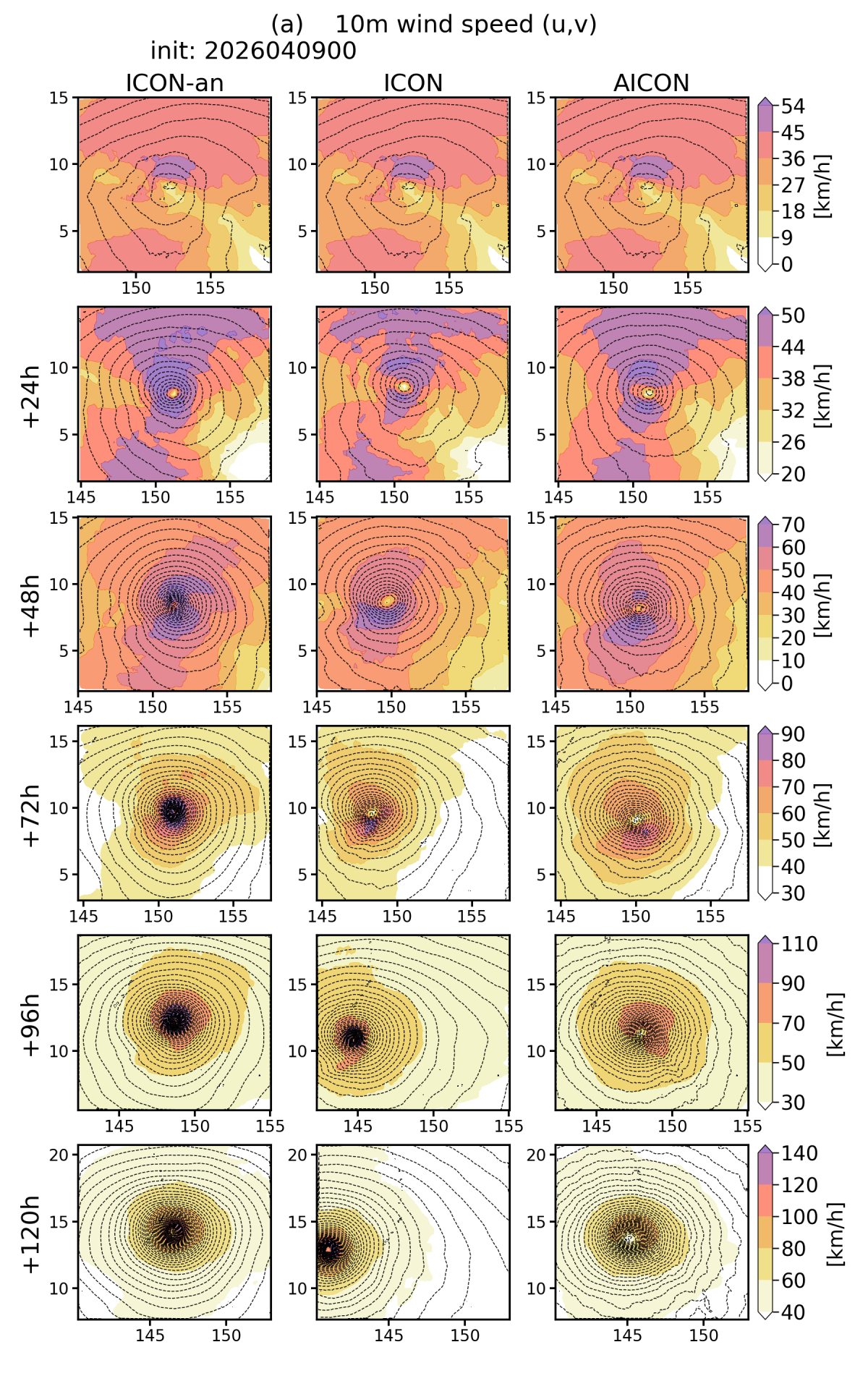}
    \includegraphics[width=0.49\linewidth]{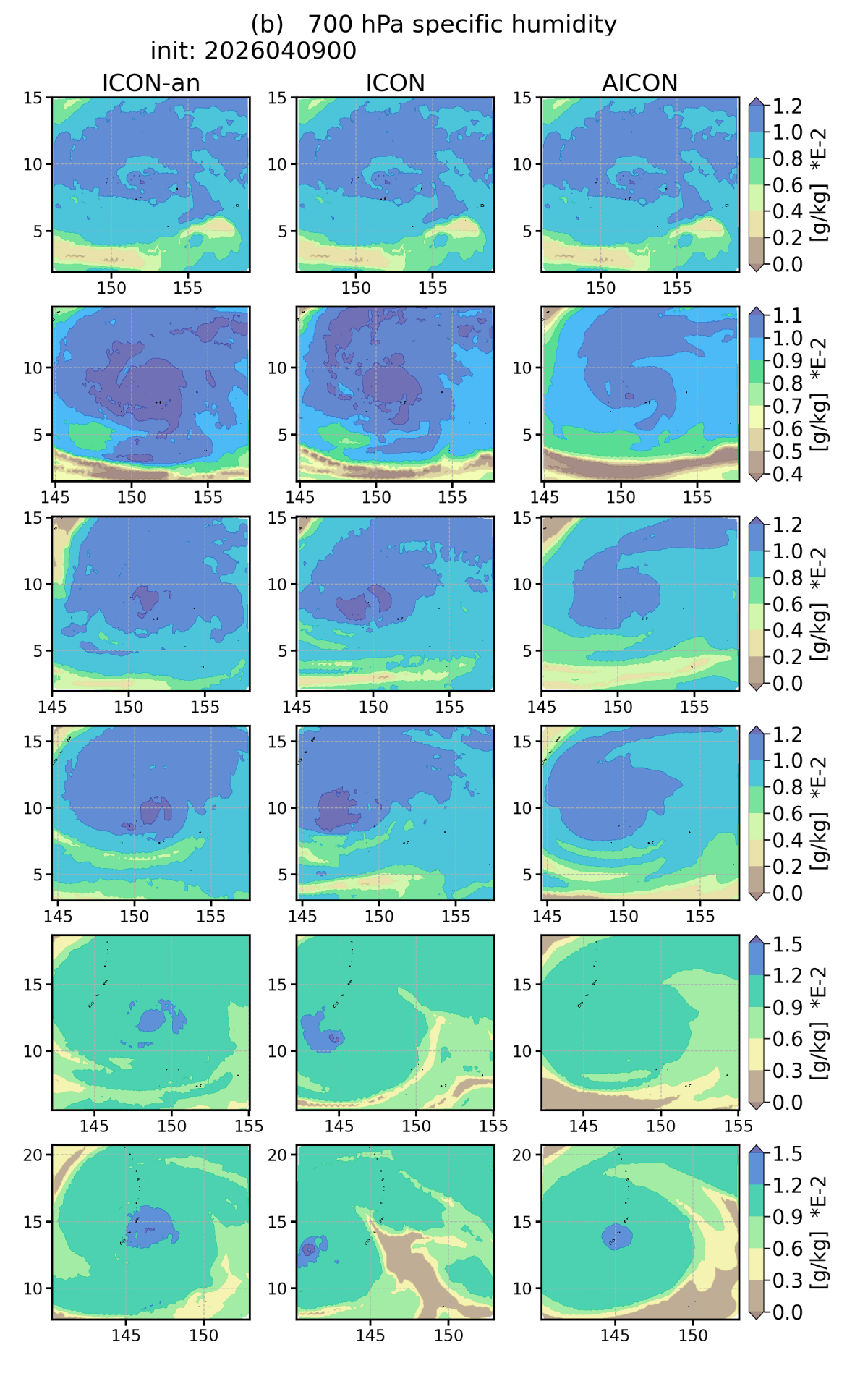}
    \caption{Intensification of Typhoon Sinlaku. The forecast initializes on 9th April 2026 00 UTC and is shown every 24 hours up to 5 days lead time. The domains are centered on the ICON analysis (ICON-an, left columns). The middle and the right columns show the ICON and AICON forecasts, respectively. The left group of columns gives 10 m wind speeds (u,v) with MSLP contours (black lines, 1 hPa spacing) (a) and the right group shows 700 hPa specific humidity (b).}
    \label{fig:SINLAKU_cyclone}
\end{figure}

Tropical cyclones are the most extreme Weather phenomena and cause huge damage by wind speeds of more than 200km/h or flooding. We compare forecasts of the Typhoon Sinlaku - a category 5 Typhoon that moved through the West-Pacific of the northern hemisphere from 8 to 22 of April 2026. Figure \ref{fig:SINLAKU_tracks} shows four different forecasts along the cyclone track on 9 to 12 of April, each initialized at 00UTC. The gray-shaded contours show the mean sea-level pressure (MSLP) at initial time of the forecast. The best track location of the Typhoon from IBTrACS 
(International Best Track Archive for Climate Stewardship, \cite{knapp2010},\cite{gahtan2024})  is indicated by the black lines. The time-colored track is the operational ICON NWP forecast covered by the ICON NWP Ensemble Prediction System (ICON-EPS) members in blue. The AICON track (red) gives better forecasts than ICON NWP on April 9 and 10, when the Typhoon is still relatively weak.

This changes on the following two days, where the Typhoon rapidly intensifies and the AICON track partly runs even outside the range of the ICON-EPS NWP members (forecast initialized on April 12). The numbers in the brackets of the panel titles give characteristic estimates of the Typhoon properties for the ICON and AICON models. From left to right the numbers in the brackets show the minimum MSLP, the maximum sustained wind speed (u,v) and the maximum gusts along the track. When comparing these numbers to those from the RSMC Tokyo – Typhoon Center (Regional Specialized Meteorological Center for Typhoons, \cite{JMA_RMSC}) and the Joint Typhoon Warning Center (JTWC, \cite{JTWC_USA}) reports given in the caption of figure \ref{fig:SINLAKU_tracks}, it becomes obvious that the ICON-NWP cyclones are not as deep as observed, but AICON cyclones are even weaker. This results in substantial underestimation of the sustained winds in AICON already when compared to ICON NWP, where the gust parametrization provides a quite reasonable estimate of the 1- and 10-minutes reported maximum wind speeds above 200km/h.

The underestimation of the cyclone intensity by AICON can also be seen in figure \ref{fig:SINLAKU_cyclone}, where the stamp maps of the forecast on April 9 show the MSLP and wind (u,v) fields (left group) as well as the specific humidity at about 700hPa (right group). In each group the left column shows the ICON operational analysis, the middle column is the ICON NWP forecast and the right column relates to AICON. The forecast panels are given every 24 h up to 5 days and the domain is centered on the ICON analysis. The left group shows that the ICON NWP cyclones drift away from the analysis during this forecast, but the intensification of the cyclone is much stronger in ICON than in AICON (right column). But AICON forecasted cyclones stay close to the analyzed track. The stamp map of specific humidity on the right of figure \ref{fig:SINLAKU_cyclone} further shows the smoothing or “regression to the mean” property of the AICON model, where the humidity fields are more uniform even during cyclone intensification.

\subsection{Subjective verification}\label{sec:subjective_evaluation}
Finally we report on the subjective verification procedure carried out by DWD's forecasters. We refrain from a detailed report on this evaluation here, since
no objective, quantitative measures were applied, that would fit the scope of this work, but we emphasize that the forecasting department was strongly involved 
in the process of bringing AICON into operations by conducting  an almost daily evaluation and inter-comparison to other models.
This complements the objective verification presented in the current section and validates the particular training strategy laid out 
in section \ref{sec:model}, in particular the decision to not use rollout and instead focus on small scale fidelity.\\

The goal of this evaluation period was to give forecasters enough time to learn the models properties, to build trust in this new kind of models and, in return, collect and use their feedback as input for further model development. 
Naturally the German forecasting section had a focus on forecast performance for Germany and Central Europe. They mainly evaluated forecast with up 
to three days lead time for the surface parameters precipitation, temperature, wind and mean sea level pressure. 
The general feedback was that the forecast quality of AICON is very similar to ICON for these variables and that the model is useful in daily forecasting practice.
AICON precipitation forecasts were perceived as of equal quality as ICON. This is for the position of rain fields as for the quantity of rain. In a few more extreme situations the precipitation maxima were under-estimated, but that also happened to be the case for other models considered in the evaluation. In general AICON was often able to reproduce local
precipitation maxima also those related to orographic structures. But also a tendency to focus on orographic influence too much was observed. Frontal
precipitation was experienced to be better represented than convective events. Considering wind speed, AICON was found to have a tendency to under predict strong wind
situations. 
Also tropical cyclones have a tendency to be on weak side, as already discussed in section \ref{sec:hurricane_tracks}. So beside the performance being
often similar to the conventional ICON model, a tendency to under predict extremes and to produce smoother fields than e.g. ICON was noted. The latter
is of course consistent with the results from the spectral analysis in section \ref{sec:spectra}.

\section{Summary}\label{sec:summary}

In this paper, we have presented AICON, the first global machine learning weather prediction (MLWP) model trained on the ICON-DREAM global reanalysis dataset \cite{valmassoi2025} and operationalized at the German national meteorological service (DWD). The high-resolution spatial 13~km and temporal 3~h ICON-DREAM reanalysis dataset differs from typical global resolution datasets of 30-50~km resolution (ERA5 \cite{hersbach2020era5}, MERRA-2 \cite{gelaro2017modern} or CRA \cite{liu2023cra}). The introduction of AICON represents a significant step in the integration of data-driven models into conventional operational
forecasting suites at DWD. Built upon the Anemoi framework \cite{aifsblogAnemoi}, the model employs a graph neural network architecture with an encoder--processor--decoder structure that respects the native icosahedral grid topology of the ICON model.

Our training strategy utilized transfer learning across multiple resolutions (52~km, 26~km, and 13~km) to ensure computational efficiency and stable convergence. A key design decision was the omission of multi-step rollout during the training phase. As demonstrated in the evaluation (Section~~\ref{sec:evaluation}), this approach prioritizes small-scale fidelity and short-range accuracy over long-range RMSE optimization. Consequently, AICON exhibits forecast variability that is closer to the conventional system on short lead times compared to rollout trained variants, though spectral analysis confirms a general damping of mesoscale activity relative to the physics-based ICON reference.

Verification against observations over a one-year period shows that AICON performs competitively with the operational ICON NWP system, particularly for surface variables in the short to medium range. The model demonstrates superior forecast consistency, evidenced by a lower flip-flop index, which is beneficial for operational forecasters. While precipitation skills are comparable to ICON in terms of Fraction Skill Score and Equitable Thread Score, we identified specific limitations, such as a tendency to underestimate extreme intensities in tropical cyclones and the emergence of spurious precipitation patterns near coastlines at long lead times. These findings highlight the inherent trade-offs in MLWP model design between statistical optimality and physical realism.

The successful operationalization of AICON at DWD in a two-step approach with technical operations since September 2025 and full operational use since March 2026 underscores the maturity of MLWP technology. It validates the hypothesis that models trained on high-resolution, non-hydrostatic reanalyses can complement conventional NWP systems in a daily forecasting routine. Future work will focus on addressing the identified limitations regarding extreme events and spectral activity, potentially through hybrid loss functions or ensemble approaches. Ultimately, AICON serves as a proof of concept that machine learning models can be robustly integrated into the critical infrastructure of weather warning systems, paving the way for hybrid forecasting paradigms that combine the speed of AI with the physical consistency of numerical models.

\section{Author contributions}
All authors contributed to writing and reviewing the manuscript.\\
TG led the design of the training curriculum and the execution of training runs, and led the drafting and revision of the manuscript.\\
MJ curated the training datasets, contributed to the software architecture and developed the automated container-integration workflow.\\
FP developed the AICON graph-construction algorithm and the operational inference software.\\
FF conducted the observation-based verification and their interpretation.\\
SW and BS conducted the spectral analyses and their interpretation.\\
SU and HR led contributions to operationalization and prepared and conducted forecast experiments.\\
RP and JK conceptualized and designed the study and provided project leadership and review.\\
AV produced the ICON-DREAM reanalysis used in this study.\\
MD conducted the hurricane-tracking evaluation.

\section{Code availability}
AICON described in this paper uses the following Anemoi package versions: \texttt{anemoi-datasets} 0.5.24, \texttt{anemoi-graphs} 0.6.0, \texttt{anemoi-models} 0.5.0, \texttt{anemoi-training} 0.4.0, \texttt{anemoi-transform} 0.1.11 and \texttt{anemoi-utils} 0.4.23.

\section*{Acknowledgments}
We would like to thank our colleagues from the forecasting department for their work on the subjective evaluation. Also, we thank our colleagues from 
the Research and Development department for their support and feedback. We have utilized several large language models (including ChatGPT-5, Claude 4.6, Gemini 3, and Mistral 3) as writing assistants, primarily to refine our text drafts. 
All ideas, analyses, and conclusions presented in this work are the sole responsibility of the authors. Any AI-suggested edits were carefully reviewed and validated by the authors before their inclusion in the final manuscript.

\vspace*{2cm}

\bibliographystyle{alpha}
\bibliography{literature}

\appendix

\section{Verification against observations (absolute scores)}\label{sec:app_abs_scores}

\begin{figure}
    \centering
    \includegraphics[width=1\linewidth]{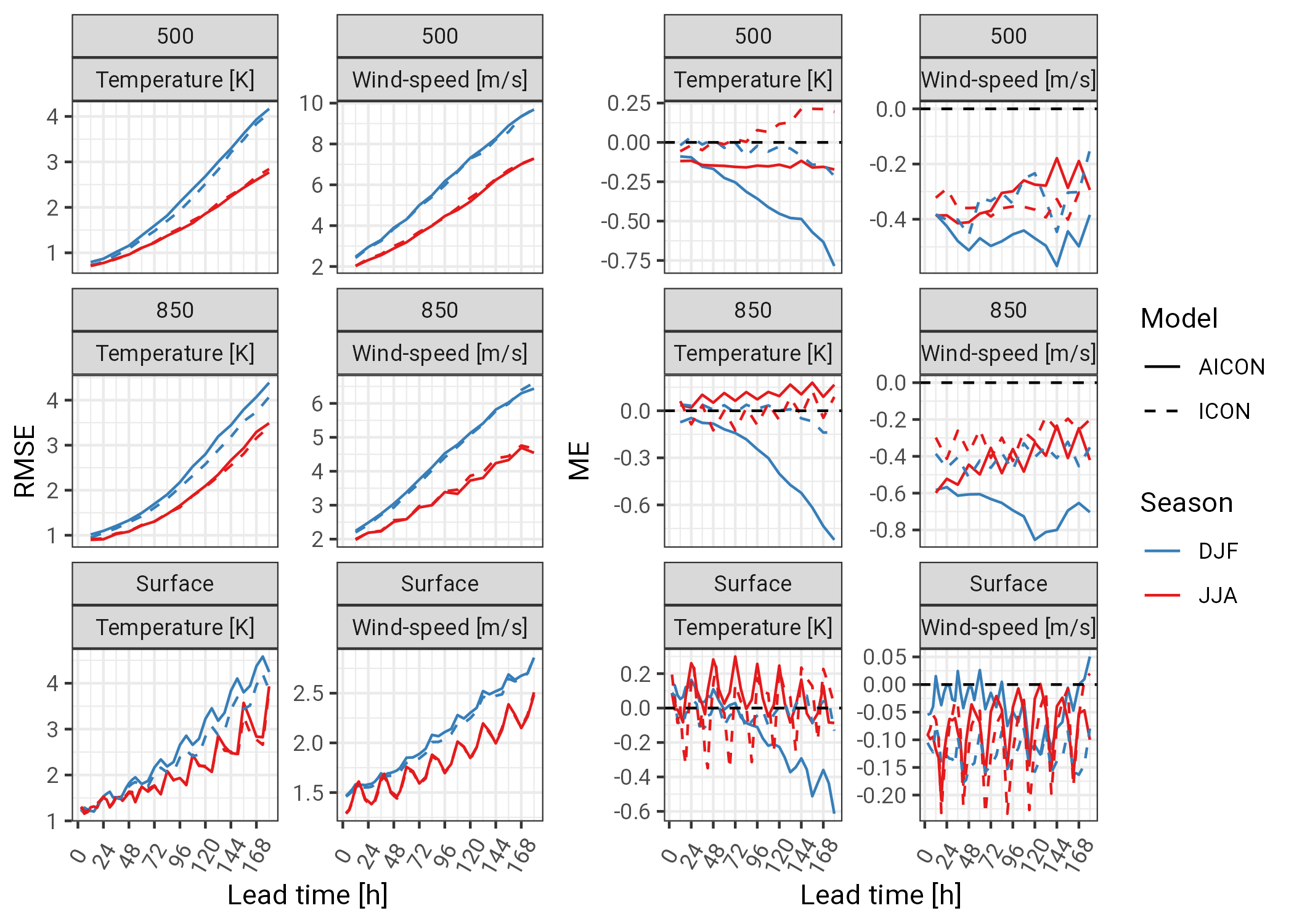}
    \caption{RMSE (left) and ME (right) against surface and radiosonde observations on the 850 hPa and 500 hPa levels in Europe. Red lines are fore the summer, blue for the winter season of 2025. Solid lines are for 00 UTC forecast runs from AICON, dashed ICON.}
    \label{fig:appendix_veri_synop_rmse}
\end{figure}

In addition to the summary score cards that show differences in scores w.r.t. the reference model ICON this section is providing some selected absolute scores. The scores are given for a winter and a summer season and are aggregated over Europe to highlight the daily cycle of errors. Figure \ref{fig:appendix_veri_synop_rmse} shows the RMSE and ME (mean error) for 2~m , 850~hPa and 500~hPa temperature and wind speed for AICON and ICON. The RMSE during summer is generally lower than in winter for the selected parameters. AICON performs similar and better than ICON during summer, while the RMSE are generally worse for AICON during winter. The winter temperature bias of AICON grows with lead time. Surface daily cycle errors are smaller for AICON in both seasons.

\section{Spectral evaluation of rollout fine-tuning}\label{sec:roll_out_spectra}

\begin{figure}
    \centering
    \includegraphics[width=1.0\linewidth]{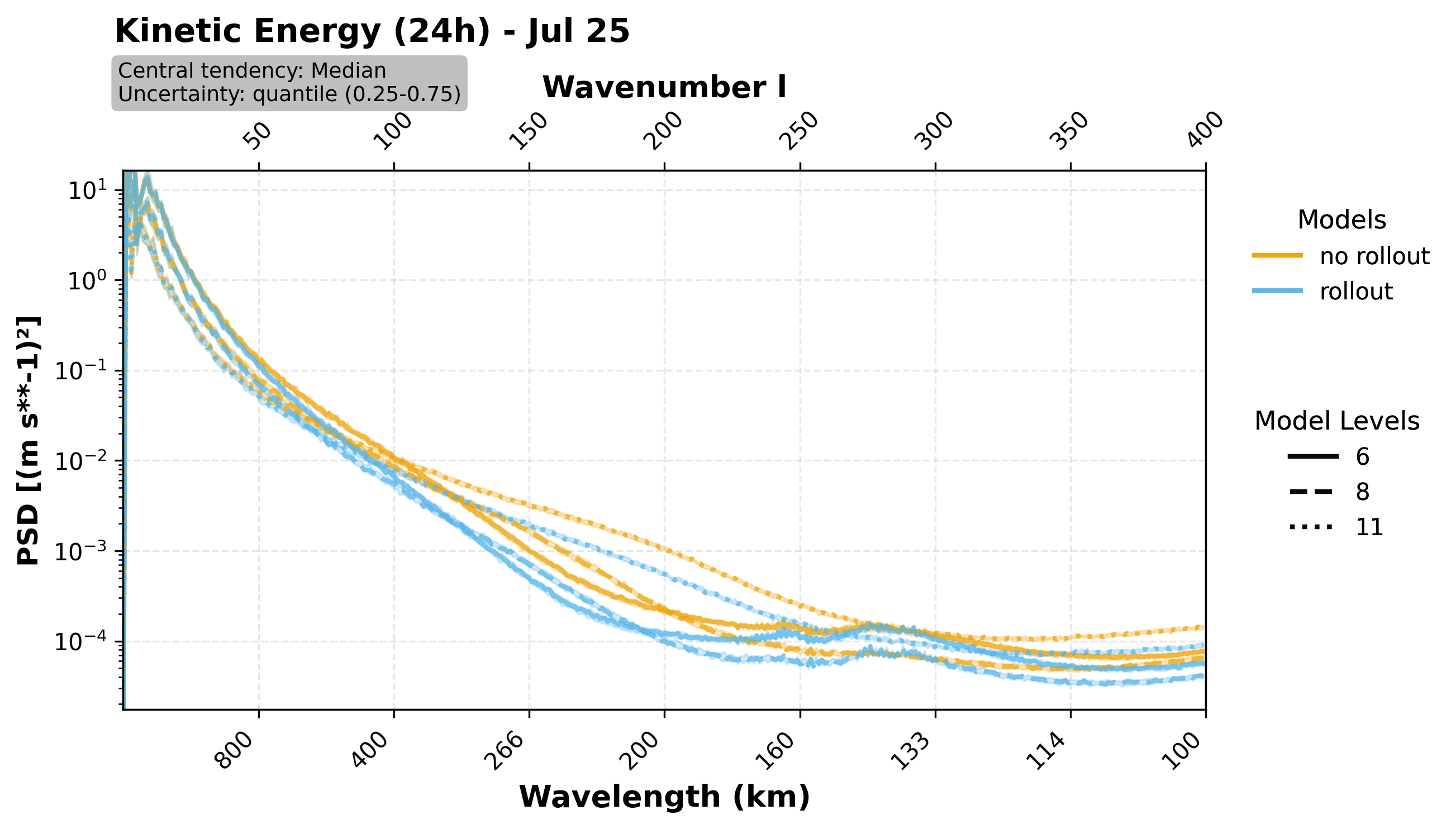}
    \caption{Kinetic Energy Spectra for AICON 24 hour forecasts for July 2025 on R03B05 grid trained with and without rollout.}
    \label{fig:aicon_r03b05_rollout_ekin_spec_full_24h}
\end{figure}

\begin{figure}
    \centering
    \includegraphics[width=1.0\linewidth]{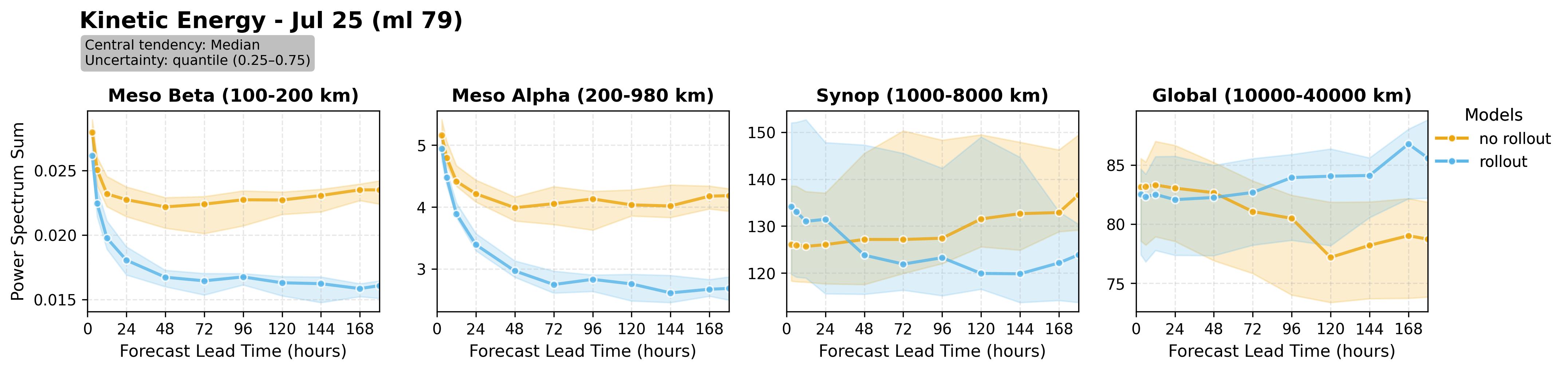}
    \caption{Kinetic Energy (model level 6/79, $\sim$ 300 hPa) activity aggregated by scales over lead time (3h - 168h) for AICON forecasts for July 2025 on R03B05 grid trained with and without rollout.}
    \label{fig:aicon_r03b05_rollout_ekin_ml6_spec_scales}
\end{figure}

Figure \ref{fig:aicon_r03b05_rollout_ekin_spec_full_24h} shows kinetic energy spectra of 24 hour forecasts (only 00:00 UTC initializations) valid in July 2025 of AICON trained with and without rollout computed as explained in section \ref{sec:spectra}. They show higher spatial variability (activity) of kinetic energy on scales between $\sim 800-150$ km for the model version without rollout. As discussed in \cite{selz2025effective} a longer rollout period during training results in smoother variable fields. Figure \ref{fig:aicon_r03b05_rollout_ekin_ml6_spec_scales} shows activity over lead time aggregated over different spectral scales. AICON without rollout (yellow) shows higher activity compared to the rollout version (blue) across all lead times on mesoscales and for lead times $\geq$ 48 h on synoptic scales. The rollout version shows a rapid decrease in activity over the first 24 hours (rollout period), flattening with higher lead times in accordance with \cite{selz2025effective}. 

\section{Verification against analysis (at observation locations)}\label{sec:app_ana_ver}

Verification against observation requires interpolation of forecast to the location of observations. In case of upper air observations (e.g. from radiosondes) this includes also a vertical interpolation. This interpolation can introduce additional errors, especially if the model levels are not close to the observation levels. When comparing models with different number of vertical levels the observed differences might be in parts due to different interpolation artifacts. Both is the case for AICON when compared to ICON w.r.t. radiosonde observations. Radiosonde reports are available and used on standard pressure levels whereas AICON and ICON provide output on model levels. With AICON having only a subset of 13 model levels from ICON the error introduced by vertical interpolation is potentially bigger for AICON than for ICON. One way to work around this issue is to perform a verification against own analysis, here we show results of a verification against own analysis at observation locations. Actually the analysis of AICON and ICON are identical, for AICON however with only 13 vertical levels. Like that the interpolation errors for forecasts and analysis are the same and factored out of the comparison of AICON and ICON. For operational reasons this could only be done for a recent period of May and June 2026.

\begin{figure}
    \centering
    \includegraphics[width=1\linewidth]{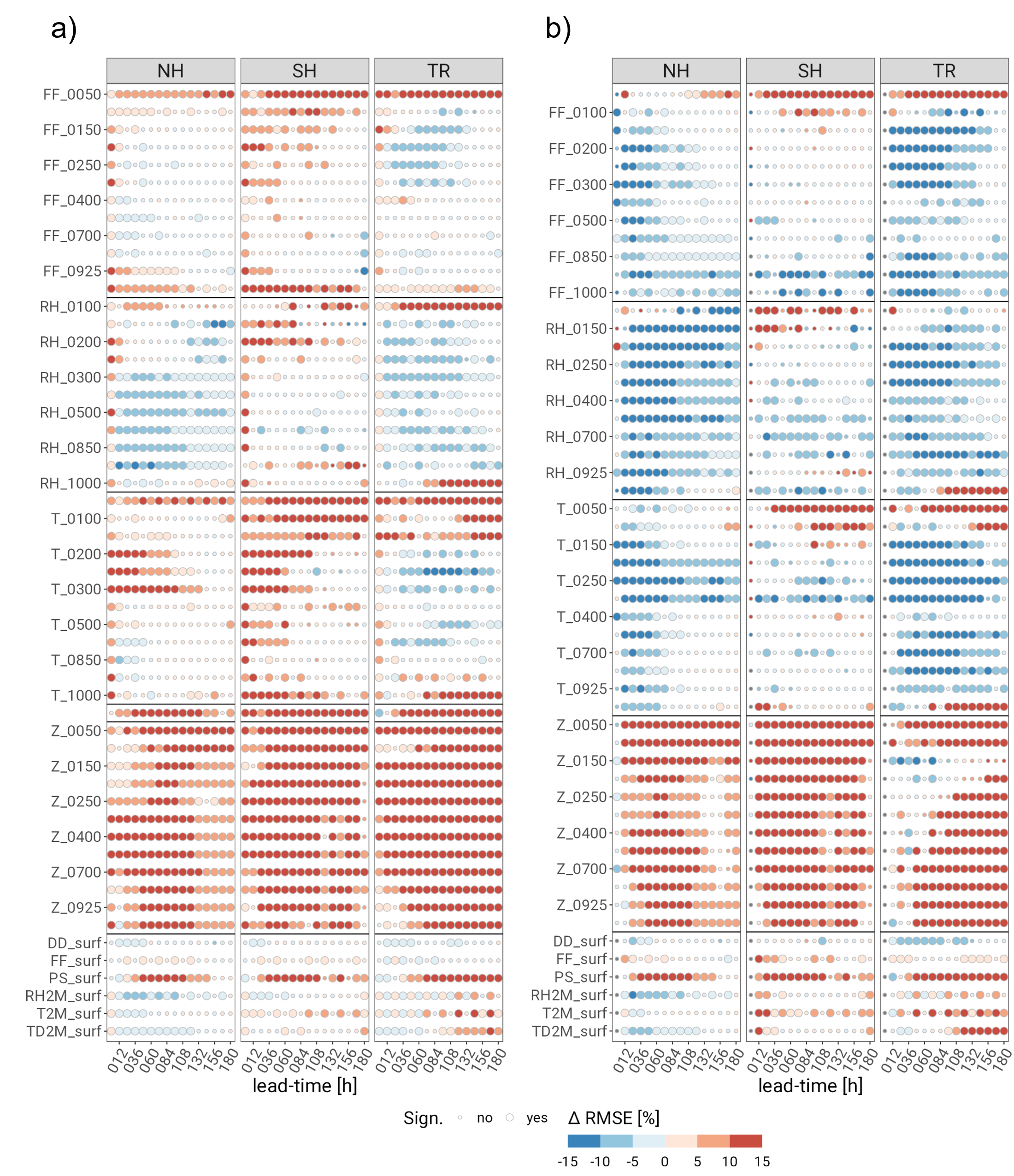}
    \caption{Verification against observations vs own analysis. Scorecard of AICON vs. ICON RMSE for forecasts from May and June 2026 subdivided in northern/southern extra tropics (NH/SH) and tropics (TR) . Wind speed (FF), relative humidity (RH), temperature (T) and geopotential height (Z) are evaluated on different pressure levels, wind direction (DD), 10~m wind speed (FF), surface pressure (PS), 2~m relative humidity, dew point temperature and temperature (RH2M, TD2M and T2M) are evaluated at surface level. The verification is based on observations from radiosondes and surface measurement sites (a) and on own analysis at observation sites (b). Where vertical interpolation is required, mostly upper air, the type of verification makes a notable difference. Here AICON shows a more pronounced improvement over ICON.}
    \label{fig:score_obs_vs_ana}
\end{figure}

The score cards in Figure \ref{fig:score_obs_vs_ana} show that the impact of switching from observation to analysis verification is significant. While the surface scores are impacted only marginally, the upper-air scores give a much more positive impression of the AICON forecast quality. AICON wind, temperature and humidity RMSE show improvements of 20\% and more over ICON when based on analysis. When based on observation this advantage shrinks to about 10\% at some levels/regions even negative. AICON scores of geopotential are problematic no matter what type of verification is chosen, still the RMSE scores bases on analysis are improved. In case of temperature it appears that the effect is especially strong in the upper troposphere, a possible explanation is that here the AICON model levels are placed furthest away from the standard pressure levels.

In conclusion, when comparing AICON to other models with a higher number of vertical levels whenever vertical interpolation to observation might play a role it is recommended to perform a verification against own analysis.

\begin{table}[p]
\centering
\small
\caption{Structural and topological specification of AICON graphs.}
\label{tab:aicon_graph_specs}
\begin{tabular}{llccc}
\toprule
\textbf{Component} & \textbf{Element / Graph Type} & \textbf{Count} & \textbf{Source Degree} & \textbf{Target Degree} \\
\midrule

% ==================== LOW RESOLUTION ====================
\multicolumn{5}{c}{\textbf{\large R03B05 (52 km)}} \\
\midrule
Nodes & Data Input Nodes & $184\,320$ & & \\
      & Hidden Mesh Nodes & $23\,042$ & & \\
\cmidrule(lr){1-3}
Edges & Encoder (Data $\rightarrow$ Hidden) & $552\,960$ & $3$ & $24$ \\
       & Processor (Hidden $\rightarrow$ Hidden) & $184\,140$ & $6-25$ & $6-25$ \\
       & Decoder (Hidden $\rightarrow$ Data) & $552\,960$ & $24$ & $3$ \\
\midrule

% ==================== MEDIUM RESOLUTION ====================
\multicolumn{5}{c}{\textbf{\large R03B06 (26 km)}} \\
\midrule
Nodes & Data Input Nodes & $737\,280$ & & \\
      & Hidden Mesh Nodes & $92\,162$ & & \\
\cmidrule(lr){1-3}
Edges & Encoder (Data $\rightarrow$ Hidden) & $2.21184\times 10^6$ & $3$ & $24$ \\
       & Processor (Hidden $\rightarrow$ Hidden) & $737\,100$ & $6-30$ & $6-30$ \\
       & Decoder (Hidden $\rightarrow$ Data) & $2.21184\times 10^6$ & $24$ & $3$ \\
\midrule

% ==================== HIGH RESOLUTION ====================
\multicolumn{5}{c}{\textbf{\large R03B07 (13 km)}} \\
\midrule
Nodes & Data Input Nodes & $2.94912\times 10^6$ & & \\
      & Hidden Mesh Nodes & $368\,642$ & & \\
\cmidrule(lr){1-3}
Edges & Encoder (Data $\rightarrow$ Hidden) & $8.84736\times 10^6$ & $3$ & $24$ \\
       & Processor (Hidden $\rightarrow$ Hidden) & $2.94894\times 10^6$ & $6-36$ & $6-36$ \\
       & Decoder (Hidden $\rightarrow$ Data) & $8.84736\times 10^6$ & $24$ & $3$ \\
\bottomrule
\end{tabular}
\end{table}

\begin{table}[p]
\centering
\small
\caption{Architectural configuration of AICON. Separated in encoder, processor and decoder}
\label{tab:aicon_model_specs}
\begin{tabular}{llc}
\toprule
\textbf{category} & \textbf{parameter} & \textbf{Value} \\
\midrule

% % ==================== Hidden / input nodes ====================
\multicolumn{3}{c}{\textbf{\large Input embedding }} \\
\midrule
feature dimension      & node input dimension   &   $2\times 93$ \\
                       & node output dimension   &   $79$ \\
                       & edge input dimension   & $2$ \\
                       & trainable node features & $8$ \\
                       & trainable edge features & $8$ \\
                       & hidden dimension   &   $512$ \\
\midrule
% ==================== ENCODER GRAPH ATTENTION ====================
\multicolumn{3}{c}{\textbf{\large Encoder }} \\
\midrule
architecture             & number of attention Heads & $16$ \\
                         & number of layers          & $1$ \\
                         & MLP hidden dimension      & $2048$ \\
\addlinespace
trainable parameters    & layer parameters       & $3\,530\,752$ \\
                        &  edge features         & $70\,778\,880$ \\
\addlinespace
distributed Execution    & number of chunks         & $2$ \\

\midrule

% ==================== PROCESSOR GRAPH ATTENTION ====================
\multicolumn{3}{c}{\textbf{\large Processor} } \\
\midrule
architecture             & number of attention heads & 16 \\
                         & Number of Layers          & 16 \\
\addlinespace
trainable parameters     & layer parameters      & $54\,738\,944$ \\
                         &  edge features       & $23\,591\,520$ \\

\midrule

% ==================== DECODER GRAPH ATTENTION ====================
\multicolumn{3}{c}{\textbf{\large Decoder}} \\
\midrule
architecture             & number of attention heads & $16$ \\
                         & number of layers          & $1$ \\
\addlinespace
trainable parameters     & layer parameters      & $3\,565\,647$ \\
                         &  edge features         & $70\,778\,880$ \\
\addlinespace
distributed Execution    & number of chunks       & $2$ \\
\bottomrule
\end{tabular}
\end{table}

\begin{table}[p]
  \caption{Model variables in training.}\label{tab:training_vaiables}
  \begin{center}
    \begin{tabular}[c]{c|c|c|c|c|c}
      \hline
      \textbf{symbol} & 
      \textbf{short name} & 
      \textbf{long name} &
      \textbf{units}  &
      \textbf{levels}  &
      \textbf{type} \\
      \hline
      $U$ & \verb|U| & zonal wind & m/s & 3D (13 levels) & prognostic \\
      $V$ & \verb|V| & meridional wind & m/s & 3D (13 levels) & prognostic \\
      $p$ & \verb|P| & pressure & Pa & 3D (13 levels) & prognostic  \\
      $q_\text{v}$ & \verb|QV| & specific humidity & kg/kg & 3D (13 levels) & prognostic \\
      $T$ & \verb|T| & temperature & K & 3D (13 levels) & prognostic \\
      \hline 
      $U_\text{10M}$ & \verb|U_10M| & zonal 10-meter wind & m/s & single &prognostic \\
      $V_\text{10M}$ & \verb|V_10M| & meridional 10-meter wind & m/s & single & prognostic \\
      $T_\text{2M}$ & \verb|T_2M| & 2-meter temperature & K & single & prognostic \\
      $RH_\text{2M}$ & \verb|RH_2M| & 2-meter relative humidity & \% & single & prognostic \\
      \hline 
      $T_\text{G}$ & \verb|T_G| & surface temperature & K & surface & prognostic \\
      $q_\text{v,s}$ & \verb|QV_S| & surface specific humidity & kg/kg & surface & prognostic \\
      $p_\text{sfc}$ & \verb|PS| & surface pressure & Pa & surface & prognostic \\
      $\alpha_{prog}$? & \verb|ALB_RAD| & surface albedo & $\in [0,1]$ & surface & prognostic \\
      $h_\text{snow}$? & \verb|H_SNOW| & snow layer depth & m & surface & prognostic \\
      \hline
      $SMI$ & \verb|SMI| & soil-moisture index & $\in [-1,2]$ & 2 soil levels & prognostic \\
      $T_\text{SO}$ & \verb|T_SO| & soil temperature & K & 2 soil levels & prognostic \\
      \hline
      $H_\text{sfc}$ & \verb|HSURF| & \makecell{surface height\\above MSL} & m & single & constant input \\
      $f_\text{land}$ & \verb|FR_LAND| & land fraction & $\in [0,1]$ & single &constant input \\
      $f_\text{lake}$ & \verb|FR_LAKE| & lake fraction & $\in [0,1]$ & single &constant input \\
      $Z_\text{0}$ & \verb|Z0| & surface roughness & m & single &constant input \\
      $\sigma_\text{SSO}$ & \verb|SSO_STDH| & \makecell{standard deviation of\\ sub-grid orography} & m & single &constant input \\
      $\epsilon_\text{sfc}$ & \verb|EMIS_RAD| & surface emissivity & $\in [-1,1]$ & single & constant input \\
      $\cos(\phi)$ &  \verb|cos_latitude| &{cosine of latitude} & $\in [-1,1]$ & single & constant input \\
      $\sin(\phi)$ &  \verb|sin_latitude| & {sine of latitude} & $\in [-1,1]$ & single & constant input \\
      $\cos(\lambda)$ & \verb|cos_longitude|& {cosine of longitude} & $\in [-1,1]$ & single & constant input \\
      $\sin(\lambda)$ & \verb|sin_longitude|& {sine of longitude} & $\in [-1,1]$  & single & constant input \\
      \hline 
      $\cos(tod)$ & \verb|cos_julian_day|&{cosine of time of day} & $\in [-1,1]$  & single & time-dependent input \\
      $\sin(tod)$ & \verb|sin_julian_day|&{sine of time of day} & $\in [-1,1]$  & single & time-dependent input \\
      $\cos(doy)$ & \verb|cos_local_time|&{cosine of day of year} & $\in [-1,1]$  & single & time-dependent input \\
      $\sin(doy)$ & \verb|sin_local_time|&{sine of day of year} & $\in [-1,1]$  & single & time-dependent input \\
      $I$ & \verb|insolation|&{insolation} & $\in [0,1]$  & single & time-dependent input \\
      \hline 
      $TP$ & \verb|TOT_PREC| & 3h accum. precipitation & mm/h & single & diagnostic \\
      \hline
    \end{tabular}
  \end{center}
\end{table}

\end{document}